\documentclass[journal]{IEEEtran}
\usepackage{amsmath,amsfonts,amsthm}
\usepackage{array}
\usepackage[caption=false,font=normalsize,labelfont=sf,textfont=sf]{subfig}
\usepackage{textcomp}
\usepackage{stfloats}
\usepackage{url}
\usepackage{hyperref} 
\usepackage{verbatim}
\usepackage{graphicx}
\usepackage{xcolor}
\usepackage{nomencl}
\usepackage{etoolbox}
\usepackage{ifthen}
\usepackage{booktabs}
\usepackage[ruled]{algorithm2e}
\usepackage{placeins}
\usepackage{multirow}
\usepackage{makecell}
\usepackage{enumitem}
\usepackage{pdfpages}

\usepackage{threeparttable} 
\usepackage{pifont} 

\allowdisplaybreaks 
\usepackage[subtle,tracking=normal]{savetrees}
\makenomenclature
\usepackage{cite}
\hypersetup{
    colorlinks=true, 
    linkcolor=black, 
    filecolor=black, 
    urlcolor=black, 
    citecolor=black, 
    linkbordercolor={1 1 1}, 
    pdfborder={0 0 0} 
}

\def\BibTeX{{\rm B\kern-.05em{\sc i\kern-.025em b}\kern-.08em
    T\kern-.1667em\lower.7ex\hbox{E}\kern-.125emX}}
\usepackage{balance}

\newtheorem*{remark}{Remark}


\newcommand{\colorblack}[1]{\textcolor{black!80!black}{#1}}
\newcommand{\bm}{\boldsymbol}

\newcommand{\RowTwo}[1]{\makecell{#1}}

\begin{document}
\bstctlcite{IEEEexample:BSTcontrol}

\newpage

\title{
   A Weighted Predict-and-Optimize Framework \\for Power System Operation Considering \\Varying Impacts of Uncertainty 
}

\author{
\IEEEauthorblockN{Yingrui Zhuang, \textit{Graduate Student Member, IEEE}}, 
\IEEEauthorblockN{Lin Cheng, \textit{Senior Member, IEEE}},
\IEEEauthorblockN{Can Wan, \textit{Senior Member, IEEE}},
\IEEEauthorblockN{Rui Xie, \textit{Member, IEEE}},
\IEEEauthorblockN{Ning Qi, \textit{Member, IEEE}},
\IEEEauthorblockN{Yue Chen, \textit{Senior Member, IEEE}}

\thanks{
This work is supported in part by National Natural Science Foundation of China (No. 52037006), 
and China Postdoctoral Science Foundation special funded project (No. 2023TQ0169).
(\textit{Corresponding author}: Yue Chen.)

Yingrui Zhuang and Lin Cheng are with the Department of Electrical Engineering, 
Tsinghua University, Beijing 100084, China (e-mail: zyr21@mails.tsinghua.edu.cn).

Can Wan is with the College of Electrical Engineering, Zhejiang University, Hangzhou 310027, China (e-mail: canwan@zju.edu.cn).

Ning Qi is with the Department of Earth and Environmental Engineering, Columbia University, NY 10027, USA (e-mail: nq2176@columbia.edu). 

Yue Chen and Rui Xie are with the Department of Mechanical and Automation
Engineering, The Chinese University of Hong Kong, Hong Kong SAR, China (e-mail: yuechen@mae.cuhk.edu.hk, ruixie@cuhk.edu.hk).}
 
}

\markboth{IEEE TRANSACTIONS ON Power Systems,~Vol.~X, No.~X, XX Month~2025}
{How to Use the IEEEtran \LaTeX \ Templates}

\maketitle

\begin{abstract}
    Prediction deviations of different uncertainties have varying impacts on downstream decision-making. 
    Improving the prediction accuracy of critical uncertainties with significant impacts on decision-making quality yields better optimization results.
    Motivated by this observation, this paper proposes a novel weighted predict-and-optimize (WPO) framework for decision-making under multiple uncertainties.
    Specifically, we incorporate an uncertainty-aware weighting mechanism into the predictive model 
    to capture the relative impact of each uncertainty on specific optimization tasks,
    and introduce a problem-driven prediction loss (PDPL) to quantify the suboptimality of the weighted predictions relative to perfect predictions in downstream optimization.
    By optimizing the uncertainty weights to minimize the PDPL, the proposed WPO framework enables adaptive assessment of uncertainty impacts and joint learning of prediction and optimization.
    Furthermore, to facilitate weight optimization, we develop a surrogate model that establishes a direct mapping between the uncertainty weights and the PDPL, 
    where enhanced graph convolutional networks and multi-task learning are adopted for efficient surrogate model construction and training.  
    \colorblack{Numerical experiments on the modified IEEE 33-bus and 123-bus systems demonstrate that the proposed WPO framework outperforms 
    the traditional predict-then-optimize paradigm, reducing the PDPL by an average of 55\% within acceptable computational time.}
\end{abstract}

\mbox{}

\section{Introduction}\label{sec:intro}
\begin{figure}[htbp]
    \setlength{\abovecaptionskip}{-0.1cm}  
    \setlength{\belowcaptionskip}{-0.1cm}   
    \centering
    \includegraphics[width=0.95\columnwidth]{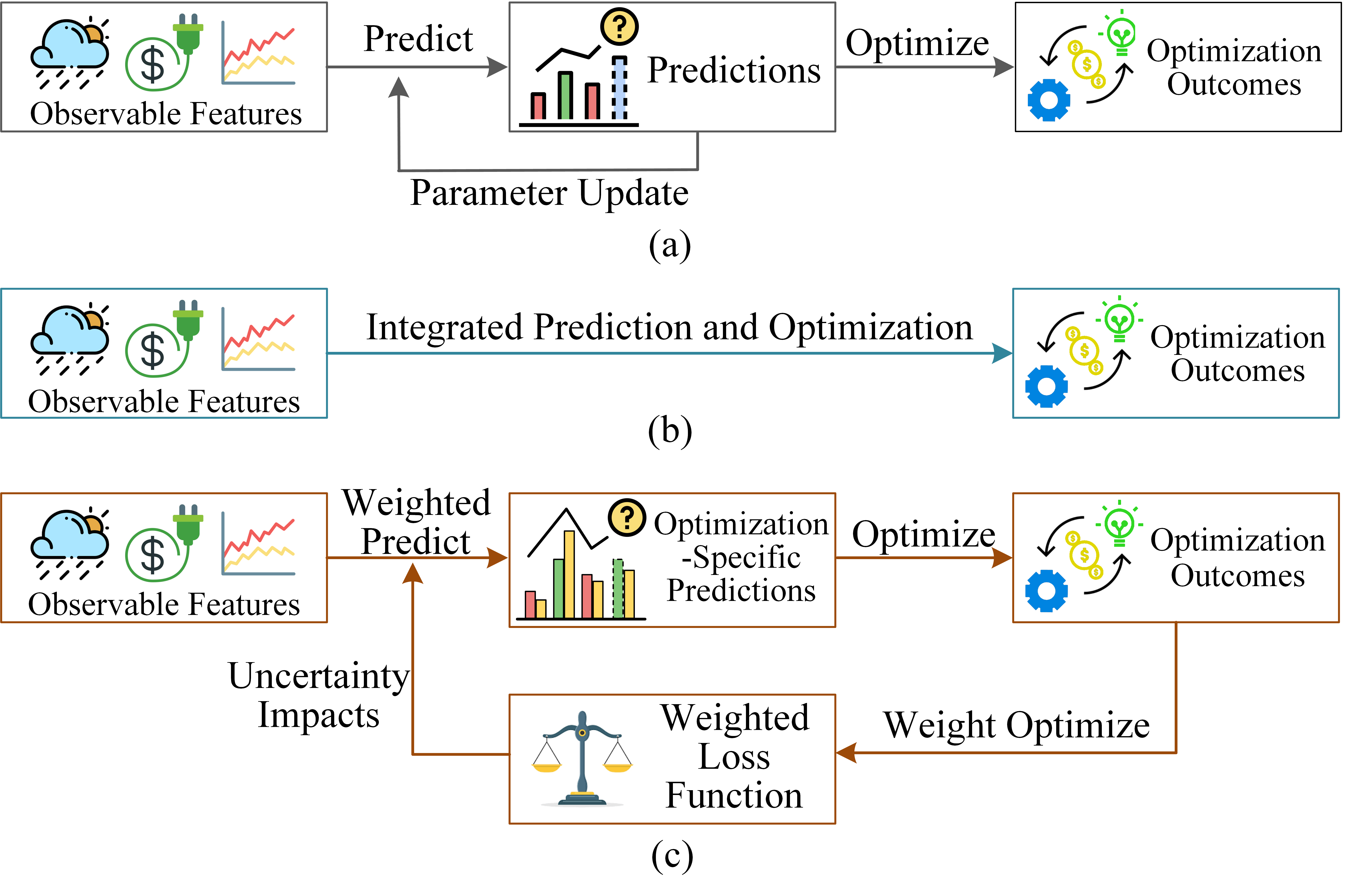}
    \caption{Diagrams of (a) traditional predict-then-optimize, (b) integrated predict-and-optimize,
    and (c) proposed weighted predict-and-optimize.}
    \label{fig:diagram}
\end{figure}

\IEEEPARstart{U}{ncertainty} brought by renewable energy sources (e.g., wind generation) and emerging types of loads (e.g., electric vehicles)
has become a significant challenge for the secure operation of power systems~\cite{Uncertaintyreview}.
Uncertainty management typically involves two key processes:
uncertainty quantification and decision-making under uncertainty.
Prediction is an effective and widely adopted method for quantifying future uncertainties 
based on observable features~\cite{predict1}. 
The resulting predictions provide essential inputs for informed decision-making under uncertainty,
forming the general predict-then-optimize paradigm, which has been extensively applied in various applications such as 
unit commitment~\cite{PTO_UC}, reserve determination~\cite{PTO_reserve}, and 
hosting capacity analysis~\cite{PTO_EVpredict}. 
Since the prediction results directly impact optimization outcomes~\cite{predicterror},
extensive studies have made substantial progress in both predictive modeling (e.g., model-driven~\cite{physicalpredict} and data-driven~\cite{datapredict}) and optimization techniques (e.g., chance-constrained optimization~\cite{qi2023}  and distributionally robust optimization~\cite{DRO}) to enhance overall decision quality.
Traditional predict-then-optimize methods generally consider prediction
and optimization as two separate and independent steps,
as illustrated in Fig.~\ref{fig:diagram}(a).
However, in practical power system operations, prediction and optimization are deeply interdependent, especially in the presence of multiple sources of uncertainty.
Different optimization tasks emphasize distinct characteristics of their predictive inputs, and the impact of prediction deviations on optimization outcomes often exhibits nonlinear, imbalanced, and problem-specific behavior~\cite{DFLreview,DFLreview1,DFLreview2}.
For instance, identical prediction deviations in load predictions impact voltage control and economic dispatch differently. 
This discrepancy arises because conventional predict-then-optimize methods neglect the task-specific requirements of downstream decision-making problems, leading to a critical limitation: the inability to generate predictions tailored to decision-specific needs, 
ultimately resulting in suboptimal decisions.

\begin{table*}[h!]
\centering
\begin{threeparttable}
\caption{Comparison with related references.}
\renewcommand{\arraystretch}{1.5}
\setlength{\tabcolsep}{0.1cm} 
\begin{tabular}{cccccc} 
    \toprule
    \RowTwo{Category} & \RowTwo{Method} & \RowTwo{Loss Function} & \RowTwo{Uncertainty\\Weight} & \RowTwo{Prediction/Optimization\\Model Requirements} & \RowTwo{Interpretability} \\
    \midrule
    \multirow{2}{*}{End-to-End} 
        & Feature $\rightarrow$ Decision 
        & \RowTwo{Decision Loss~\cite{PAO_ES}\\Decision Similarity~\cite{PAO_trading}\\Combined Decision \& Statistical Loss~\cite{Beichter}} 
        & \ding{55} 
        & General 
        & Low \\
            \addlinespace[3pt] 

        & Feature $\rightarrow$ Objective 
        & Minimal Objective~\cite{chen2024neural,PAO_costinterval} 
        & \ding{55} 
        & General 
        & Low \\ 
            \addlinespace[6pt] 

    \multirow{3}{*}{\RowTwo{Differentiable\\[2pt]Embedding}} 
        & Prediction-Embedded 
        & \RowTwo{Minimal Objective~\cite{PAO_Inertia}\\Decision Loss~\cite{PAO_UC1}} 
        & \ding{55} 
        & Convex/Linear Prediction Model 
        & \RowTwo{Medium} \\
        \addlinespace[3pt] 

        & Optimization-Embedded 
        & \RowTwo{Objective~\cite{PAO_VPP1, PAO_Seqmarket}\\Decision Loss~\cite{chung2022decision}\\Combined Decision \& Statistical Loss~\cite{PAO_priceES,PAO_voltagecontrol}} 
        & \ding{55} 
        & \RowTwo{Differentiable Optimization\\[2pt]with Direct Gradient Access} 
        & \RowTwo{Medium} \\
        \addlinespace[3pt] 

        & Loss-Fitted 
        & Decision Loss~\cite{PAO_costpredict,PAO_wind,PAO_zhang2023D} 
        & \ding{55} 
        & Few Number of Uncertainties 
        & Medium \\
            \addlinespace[6pt] 

    This Paper 
        & Weight-Integrated 
        & Problem-Driven Weighted Prediction Loss 
        & $\checkmark$ 
        & General 
        & High \\
    \bottomrule
\end{tabular}
\begin{tablenotes}
    \item 
    $\checkmark$: Considered, \ding{55}: Not considered.
\end{tablenotes}
\end{threeparttable}\vspace{-0.3cm}
\end{table*}

\colorblack{
Distinct from traditional predict-then-optimize methods, 
predict-and-optimize approaches integrate prediction with downstream optimization to improve overall decision-making quality~\cite{DFLreview,DFLreview1,DFLreview2}, 
as illustrated in Fig.~\ref{fig:diagram}(b). 
Additionally, the decision loss~\cite{SPO} has been proposed to quantify
the suboptimality of decisions derived from imperfect predictions relative to the ideal decisions under perfect predictions.
In the existing literature, three general categories of predict-and-optimize methods have been developed:
\textit{(i) End-to-End} methods
establish direct mappings from observable features to optimization outcomes 
(e.g., decisions~\cite{PAO_ES,PAO_trading,Beichter} and objectives~\cite{PAO_costinterval,chen2024neural}).
However, due to the absence of explicit intermediate results, 
end-to-end methods are often questioned for insufficient interpretability and credibility in practical applications. 
Moreover, since end-to-end methods embed both complex predictive and optimization models into a single direct mapping,
they suffer from high mapping model complexity and limited generalization capabilities.
\textit{(ii) Differentiable Embedding} methods
retain the prediction process,
but simplifies the predictive or optimization model into a differentiable form
and embeds it into the other model to enable a joint learning of prediction and optimization.
This includes 
\textit{(a)} predictive model embedding~\cite{PAO_Inertia, PAO_UC1}, 
where nonlinear predictors are replaced by simpler convex or linear models (e.g., linear regression) 
and embedded into the optimization layer,
at the cost of reduced feature extraction capabilities and limited applicability to complex prediction tasks;
\textit{(b)} optimization model embedding~\cite{PAO_priceES, PAO_voltagecontrol,chung2022decision}, 
where the optimization process is encoded as a differentiable layer~\cite{optnet} within the predictive model, 
which allows gradient-based training from optimization outcomes,
but is restricted to relatively simple decision tasks. 
\textit{(c)} loss function fitting~\cite{PAO_costpredict, PAO_wind,PAO_zhang2023D},
where a mapping between statistical prediction loss and deviations in optimization objectives is fitted 
and used to replace the statistical loss function in predictive model training.
However, this approach is limited to few number of uncertainties due to the difficulty of explicitly fitting high-dimensional relationships.
Despite these advances, 
predict-and-optimize methods still face persistent challenges in interpretability,
adaptability and generalization when applied to complex prediction and decision-making contexts. 
}

In contrast to existing methods,
we propose a new perspective to enable predict-and-optimize framework by prioritizing the critical uncertainties
that have significant impacts on optimization outcomes. 
In power system operations involving multiple sources of uncertainty,
the impacts of prediction deviations of uncertainties on optimization vary significantly,
depending on the specific problem characteristics and the role each uncertainty plays within the optimization~\cite{PAO_DR}.
For instance, in voltage control problems, 
prediction deviations in load demands at nodes with lower voltage security margins have a greater impact on system security.
Therefore, improving the prediction accuracy of critical uncertainties can mitigate the adverse effects of prediction errors and enhance overall optimization performance.
Weights can be incorporated into the predictive loss function to indicate the relative importance of uncertainties. 
With all other influencing factors unchanged, 
the predictive model tends to reduce the prediction deviations of uncertainties assigned higher weights.
These weights should be optimized to accurately represent each uncertainty’s impact on the optimization outcomes.
Serving as a bridge between prediction and optimization,
the weights allow the predictive model to be tailored to specific decision-making objectives,
thereby improving both the interpretability and adaptability of the predict-and-optimize framework.
Regarding other weighting strategies,
traditional predict-then-optimize paradigms generally assign equal weights to all uncertainties.
Some studies assign weights to data samples~\cite{weight2} or combined multiple predictive models~\cite{weight3} 
for higher prediction accuracy.
However, such methods neglect the characteristics of downstream decision-making process and may consequently lead to suboptimal decisions.

In this paper, we propose a novel weighted predict-and-optimize framework
for uncertainty management in power system operations, as illustrated in Fig.~\ref{fig:diagram}(c). 
Specifically, our contributions are as follows:
\colorblack{
\begin{enumerate}
    \item \textit{Weighted Predict-and-Optimize Framework:}
    We propose a novel weighted predict-and-optimize framework for uncertainty quantification and
    management in power systems.
    This framework introduces an uncertainty-aware weighting mechanism into the loss function of the predictive model 
    to reflect the relative impact of each uncertainty on the optimization outcome.
    A problem-driven prediction loss (PDPL) is further defined to explicitly quantify the suboptimality of the weighted predictive model relative to the ideal case with perfect predictions.
    The weight settings are then optimized to minimize the PDPL through a surrogate model that captures the mapping between the weights and the PDPL,
    enabling integrated learning of prediction with optimization.
    As a result, the framework yields predictions tailored to specific decision-making tasks,
    thus enhancing interpretability and adaptability.
    \item \textit{Surrogate Model for Weight Optimization:} 
    To facilitate weight optimization for minimizing the PDPL,
    we develop a surrogate model that establishes a direct mapping between weights and PDPL,
    and optimize the weights via gradient descent.
    Specifically, leveraging the graph structure of the uncertainties distributed in power system topology,
    we construct an enhanced graph convolutional network (GCN) as the surrogate model,
    which is verified with high mapping precision compared to traditional models.
    \item \textit{Multi-Task Learning for Reducing Computational Burden:} 
    Constructing the dataset for training the surrogate model requires training numerous predictive models 
    under different weight settings, which can be computationally expensive. 
    To address this issue, we propose a multi-task learning (MTL) method,
    which enables joint learning of multiple predictive models through an information-sharing mechanism and task-specific output layers.
    MTL significantly reduces the computational burden and the training time,
    while maintaining satisfactory prediction performance.
\end{enumerate}
}

The remainder of the paper is organized as follows. 
Section~\ref{sec:PTO} introduces the problem statements of predict-then-optimize.
The methodology of the proposed WPO framework is presented in Section~\ref{sec:methodology}.
Formulation of a classical uncertainty management problem is presented in Section~\ref{sec:risk}. 
Numerical case studies are presented in Section~\ref{sec:case}. 
Finally, conclusions are summarized in Section~\ref{sec:conclusion}.

\section{Formulation of Traditional \\Predict-then-Optimize}\label{sec:PTO}
\colorblack{
In this section, we present the general formulation of traditional predict-then-optimize paradigm,
which separates prediction and optimization as two independent tasks.
}

\subsection{Prediction}
Generally, the prediction task is formulated as: 
\begin{equation}\label{eq:predict}
    \hat{\xi} = \varphi_{\theta}(s),
\end{equation}
where $\varphi_{\theta}(\cdot)$ is a predictive model parameterized by $\theta$,
\colorblack{
which can be implemented using physics-based, statistical, or artificial intelligence models,
and should be properly designed to ensure satisfactory predictive performance.
}
$s$ denotes the observable features,
\colorblack{
and generally comprises: (i) historical statistics of the target variable $\hat{\xi}$, 
(ii) historical statistics or predictive information of external explanatory variables 
(e.g., meteorological data, geographical information).}
$\hat{\xi} \in \mathbb{R}^{n\times T}$ and $\xi \in \mathbb{R}^{n\times T}$ represent the predicted results and the corresponding ground-truth realizations of $n$ uncertain variables, respectively.
\colorblack{
    $T$ denotes the prediction horizon and can be set depending on specific applications.
Note that $\hat{\xi}$ can be scalar values for deterministic predictions,
and probability distributions for probabilistic predictions.
}

\colorblack{
Given a prediction–observation pair $(\hat{\xi}, \xi)$ in the training dataset,
the loss function quantifies the prediction discrepancy across all $n$ uncertain variables:
}
\begin{equation}\label{eq:predictloss}
    \mathcal{L}(\hat{\xi},\xi) :=  \sum_{i=1}^n \frac{1}{n}\ell (\hat{\xi}_{i}, \xi_{i}),
\end{equation}
\colorblack{
where $\ell (\hat{\xi}_{i}, \xi_{i})$ is the prediction loss for the $i$-th uncertain variable.}
For deterministic prediction, mean absolute error (MAE) and mean squared error (MSE) are commonly adopted for $\ell(\cdot)$.
For probabilistic prediction, pinball loss and continuous ranked probability score (CRPS) are commonly adopted for $\ell(\cdot)$.
\colorblack{
Notably, 
if raw units of $\xi_i$ differ, a normalization procedure would be necessary to guarantee fair comparability.
}

\colorblack{
The total loss function on the training dataset $\mathcal{D}^{\mathrm{F}}$ is:
\begin{equation}\label{eq:totalloss}
     \mathcal{L}^{\mathrm{total}} := \mathbb{E}_{(s, \xi) \in \mathcal{D}^{\mathrm{F}}} [\mathcal{L}( \varphi_{\theta}(s),\xi)],
\end{equation}
where $\mathbb{E}[\cdot]$ represents the expectation operator, 
$\mathcal{D}^{\mathrm{F}} $ denotes the training dataset for prediction with $ |\mathcal{D}^{\mathrm{F}}| $ data samples. }
Given the predictive model structure and the loss function,
the objective of predictive model training is to determine  $ \theta^\ast $ 
that minimizes the total loss:
\begin{equation}\label{eq:thetaoptimization}
    \theta^\ast = \underset{\theta}{\arg\min}\;  \mathcal{L}^{\mathrm{total}} = \underset{\theta}{\arg\min}\;  \mathbb{E}_{(s, \xi) \in \mathcal{D}^{\mathrm{F}}} [\mathcal{L}( \varphi_{\theta}(s),\xi)].
\end{equation}

\subsection{Optimization}

\colorblack{
    Eqs.~\eqref{eq:predict}-\eqref{eq:thetaoptimization} complete the prediction process.
    Then, the trained predictive model $\varphi_{\theta^\ast}(\cdot)$ is adopted to generate predictions $\hat{\xi}$ for uncertainty quantification,
which serve as input parameters to
a down-stream optimization model to support decision-making under uncertainty:}
\begin{equation}\label{eq:optimization}
    z_{\hat{\xi}}^\ast = \underset{z \in Z}{\arg\min} \; f(z, \hat{\xi}),
\end{equation}
where $z\in Z$ denotes the decision variables, $Z$ represents the feasible set, and $f(\cdot)$ denotes the objective function.
Corresponding to the prediction formulation,
~\eqref{eq:optimization} can be deterministic optimization or stochastic optimization.

As indicated in~\eqref{eq:predict}-\eqref{eq:thetaoptimization} and~\eqref{eq:optimization},
traditional predict-then-optimize paradigm separates prediction and optimization as two independent steps.
Next, we introduce the proposed weighted predict-and-optimize framework.

\section{Methodology of \\Weighted Predict-and-Optimize}\label{sec:methodology}

In the traditional predict-then-optimize paradigm,
the loss function~\eqref{eq:predictloss} assigns equal weights of $1/n$ to all uncertainties,
giving them equal importance.
However, critical uncertainties with significant impacts on decision-making should be predicted more accurately to mitigate
the negative effects of prediction deviations and thereby enhance decision quality.
Thus, we incorporate variable-specific weights into the predictive model to prioritize critical uncertainties, 
and adopt a problem-driven prediction loss that explicitly quantifies the suboptimality of 
imperfect predictive model with respect to decision-making. 
Furthermore, the weights are optimized to minimize the PDPL,
thereby enabling the integration of prediction and downstream optimization.

\subsection{Weighted Predictive Model}
We incorporate variable-specific weights into the loss function, reformulating the conventional loss in~\eqref{eq:predictloss} as a weighted form:
\begin{equation}\label{eq:weightedpredictloss}
    \mathcal{L}_{\omega}(\hat{\xi}, \xi) := \sum_{i=1}^n  \omega_i \ell(\hat{\xi}_{i}, \xi_{i}),
\end{equation}
where $\omega_i$ denotes the weight assigned to the $i$-th uncertain variable, 
satisfying $0 \leq \omega_i \leq 1$ and $\sum_{i=1}^n \omega_i = 1$. 
Notably, setting $\omega_i = 1/n,\, \forall i$ recovers the conventional loss function~\eqref{eq:predictloss}.

The predictive model is subsequently trained using the weighted loss function~\eqref{eq:weightedpredictloss}:
\begin{equation}\label{eq:optimalTheta}
    \theta_{\omega}^\ast = \underset{\theta}{\arg\min}\;  \mathcal{L}^{\mathrm{total}}_{\omega} = \underset{\theta_{\omega}}{\arg\min} \;\mathbb{E}_{(s, \xi) \in \mathcal{D}^{\mathrm{F}}}
    \big[\mathcal{L}_{\omega}(\varphi_{\theta_{\omega}}(s), \xi)\big],
\end{equation}
where $\theta_{\omega}^\ast$ denotes the optimal parameters of the predictive model  trained 
under the weighted loss function~\eqref{eq:weightedpredictloss} with weights $\omega$.

\colorblack{
Generally, gradient descent is employed to optimize $\theta_{\omega}$ 
by iteratively updating $\theta_{\omega}$ in the direction of the negative gradient of 
$\mathcal{L}^{\mathrm{total}}_{\omega}$:
\begin{equation}
    \theta_{\omega}^{k+1} = \theta_{\omega}^{k} - \eta \cdot \nabla_{\theta_{\omega}} \mathcal{L}^{\mathrm{total}}_{\omega},
\end{equation}
where $\eta$ is the learning rate, $k$ is the iteration index,
and $\nabla_{\theta_{\omega}} \mathcal{L}^{\mathrm{total}}_{\omega}$ is expanded as:
\begin{equation}
    \nabla_{\theta_{\omega}} \mathcal{L}^{\mathrm{total}}_{\omega} = \mathbb{E}_{(s, \xi) \in \mathcal{D}^{\mathrm{F}}} 
    [\sum_{i=1}^n \omega_i \cdot \nabla_{\theta_{\omega}} \ell (\varphi_{\theta_{\omega}}(s)_i, \xi_{i})].
\end{equation}
}

\colorblack{
This indicates that a larger weight $\omega_i$ increases the contribution of its loss gradient $\nabla_{\theta_{\omega}}\ell\bigl(\varphi_{\theta_{\omega}}(s)_i, \xi_i\bigr)$ to the overall gradient. 
Consequently, parameter updates are primarily guided by this component’s loss. Compared with the traditional equally weighted case, uncertainties assigned higher weights than $1/n$ tend to achieve higher relative prediction accuracy.
}

\subsection{Problem-Driven Prediction Loss}
Given a prediction result $\hat{\xi}$,
the corresponding optimal decision is $z_{\hat{\xi}}^\ast$.
After the true realization $\xi$ is revealed,
the decision $z_{\hat{\xi}}^\ast$ is subsequently implemented under $\xi$.
Decision loss is employed to quantify the suboptimality of $z_{\hat{\xi}}^\ast$ derived from $\hat{\xi}$ 
relative to the optimal decision $z_{\xi}^\ast$ derived from $\xi$:
\begin{equation}\label{eq:decisionLoss}
    \mathcal{L}^{\mathrm{D}}(\hat{\xi}, \xi) := f(z_{\hat{\xi}}^\ast, \xi) - f(z_{\xi}^\ast, \xi).
\end{equation}

We have $\mathcal{L}^{\mathrm{D}}(\hat{\xi}, \xi) \geq 0$. 
Under perfect prediction, i.e., $\hat{\xi} = \xi$, we have $\mathcal{L}^{\mathrm{D}}(\hat{\xi}, \xi) = 0$.
Building upon~\eqref{eq:decisionLoss}, 
we define the problem-driven prediction loss of imperfect predictive model  $\varphi_{\theta}$
as the expected decision loss across all uncertainty realizations in $\mathcal{D}^{\mathrm{F}}$:
\begin{equation}\label{eq:expectedDecisionLoss}
    \mathcal{L}^{\mathrm{D}}_{\varphi_{\theta_{\omega}}} := 
    \mathbb{E}_{(s, \xi) \in \mathcal{D}^{\mathrm{F}}}\big[\mathcal{L}^{\mathrm{D}}(\varphi_{\theta_{\omega}}(s), \xi)\big].
\end{equation}

\colorblack{
PDPL reflects the real performance of prediction-derived strategies relative to an ideal benchmark, 
bridging statistical prediction deviations and decision-level deviations,
providing a problem-driven measure of predictive performance.
For instance, in power system scheduling, predicted load deviations may cause operators 
to adjust their strategies, resulting in extra procurement costs or curtailment losses.
}

\colorblack{
Note that the proposed WPO framework does not require the predictive and optimization models to be specific forms.
Thus, general formulations are presented.
}
But we reasonably require~\eqref{eq:decisionLoss} and~\eqref{eq:expectedDecisionLoss} to satisfy the following
three assumptions:

\begin{enumerate}[label=\alph*)]
    \item The predictive model exhibits satisfactory performance, 
          ensuring that $\hat{\xi}$ remains within an acceptable range of deviation from $\xi$.
    \item  The optimization problem~\eqref{eq:optimization} is well-defined and can be solved to optimality.
    \item Sufficient resources are available to manage potential prediction deviations, even costly. 
          Thus, there exists a feasible solution to~\eqref{eq:optimization} for any practical $\hat{\xi}$. 
\end{enumerate}

\subsection{Weight Optimization}
To achieve integrated learning of prediction and optimization,
we aim to optimize the weights $\omega$ to minimize the PDPL:
\begin{equation}\label{eq:optimalWeight}
    \begin{split}
            \omega^\ast &= \underset{\omega}{\arg\min} \;\mathcal{L}^{\mathrm{D}}_{\varphi_{\theta_{\omega}^\ast}}\\
            \text{s.t.} \ &\eqref{eq:predict},\ \eqref{eq:optimization}-\eqref{eq:expectedDecisionLoss},\\
            &0 \leq \omega_i \leq 1,\ \sum_{i=1}^n \omega_i = 1
    \end{split}
\end{equation}
where $\mathcal{L}^{\mathrm{D}}_{\varphi_{\theta_{\omega}^\ast}}$ is generated by applying $\theta=\theta_{\omega}^\ast$ to~\eqref{eq:expectedDecisionLoss}.
In~\eqref{eq:optimalWeight}, the weights $\omega$ serve as a bridge between prediction and downstream optimization.
By optimizing the weights $\omega$ to minimize the PDPL, 
we obtain weights reflecting the problem-specific relative importance of uncertainties, 
thus generating predictions tailored to the decision-making task and thereby improving decision quality.

However,
the predictive model $\varphi_{\theta_{\omega}}(\cdot)$, often implemented by a complex neural network, 
combined with the optimization terms in $\mathcal{L}^{\mathrm{D}}(\varphi_{\theta_{\omega}}(s), \xi)$, 
makes the optimization of $\omega$ in~\eqref{eq:optimalWeight} highly nonconvex,
which cannot be solved directly.
Thus, we propose a differentiable surrogate model $\phi_{\vartheta }(\cdot)$
to build a direct mapping between $\omega$ and $\mathcal{L}^{\mathrm{D}}_{\varphi_{\theta_{\omega}}}$ using a data-driven approach,
thereby facilitating the weight optimization.
The mapping process is illustrated as follows:
\begin{equation}\label{eq:surrogate}
    \phi_{\vartheta } (\omega) \rightarrow \mathcal{L}^{\mathrm{D}}_{\varphi_{\theta_{\omega}}}. 
\end{equation}

The mapping performance of $\phi_{\vartheta }(\cdot)$ can be evaluated by:
\begin{equation}\label{eq:surrogateLoss}
    \mathcal{L}^{\mathrm{S}} := \frac{1}{|\mathcal{D^{\text{S}}}|}\sum_{(\omega, \mathcal{L}^{\mathrm{D}}_{\varphi_{\theta_{\omega}}}) \in \mathcal{D^{\text{S}}}}
    (\phi_{\vartheta }(\omega) - \mathcal{L}^{\mathrm{D}}_{\varphi_{\theta_{\omega}}})^2,
\end{equation}
where $\mathcal{D}^{\text{S}}$ is the dataset for surrogate model training,
consisting of $|\mathcal{D}^{\text{S}}|$ data pairs $(\omega, \mathcal{L}^{\mathrm{D}}_{\varphi_{\theta_{\omega}}})$.

The surrogate model $\phi_{\vartheta }(\cdot)$ is trained to minimize~\eqref{eq:surrogateLoss}:
\begin{equation}\label{surrogateoptimize}
    \vartheta^\ast = \underset{\vartheta}{\arg\min} \; 
    \frac{1}{|\mathcal{D^{\text{S}}}|}\sum_{(\omega, \mathcal{L}^{\mathrm{D}}_{\varphi_{\theta_{\omega}}}) \in \mathcal{D^{\text{S}}}}
    (\phi_{\vartheta }(\omega) - \mathcal{L}^{\mathrm{D}}_{\varphi_{\theta_{\omega}}})^2.
\end{equation}

With trained $\phi_{\vartheta^\ast }(\cdot)$, we achieve accurate mapping 
from $\omega$ to $\mathcal{L}^{\mathrm{D}}_{\varphi_{\theta_{\omega}}}$.
Then, $\omega$ is optimized to minimize the PDPL 
in~\eqref{eq:optimalWeight} by performing gradient descent on $\phi_{\vartheta^\ast }(\cdot)$:
\begin{equation}\label{eq:weightUpdate}
    \omega^{\kappa+1} = \omega^{\kappa} - \alpha \frac{\partial \phi_{\vartheta^\ast }(\omega^{\kappa})}{\partial \omega^{\kappa}}, 
\end{equation} 
where $\kappa$ is the iteration step and $\alpha>0$ is the learning rate. 
The iteration process~\eqref{eq:weightUpdate} is repeated until $\phi_{\vartheta^\ast }(\omega)$ converges to its minimum, 
yielding the optimal weight setting $\omega^\ast$.

\colorblack{
    The convergence of~\eqref{eq:weightUpdate} is theoretically guaranteed if $\phi_{\vartheta^\ast }(\cdot)$ is Lipschitz smooth,
    which can be ensured through appropriate surrogate model design (e.g., neural networks with bounded activation functions) and training strategies (e.g., regularization)~\cite{gradientconverge}.
}

However, two major challenges exist in~\eqref{surrogateoptimize}:
\begin{enumerate}
    \item \textbf{Mapping ability of surrogate model:}
    $\phi_{\vartheta }(\cdot)$  must be capable of capturing the high-dimensional and non-linear relationship
    between $\omega$ and $\mathcal{L}^{\mathrm{D}}_{\varphi_{\theta_{\omega}}}$,
    and of providing reliable guidance for weight optimization  via~\eqref{eq:weightUpdate}.
    \item \textbf{Computation burden in large-scale predictive model training:}
    Training $\phi_{\vartheta }(\cdot)$ requires a sufficiently large dataset composed of $(\omega,\ \mathcal{L}^{\mathrm{D}}{\varphi{\theta_{\omega}}})$ pairs.
    Each data sample in $\mathcal{D}^{\mathrm{S}}$ necessitates a complete training cycle of the predictive model 
    $\varphi_{\theta_\omega}(\cdot)$.
    Thus, generating $\mathcal{D}^{\mathrm{S}}$ can be computationally expensive and time-consuming.
\end{enumerate}

Next, we present an enhanced graph convolutional network for surrogate model construction to address challenge 1 in Section~\ref{sec:surrogate},
and a multi-task learning method to address challenge 2 in Section~\ref{sec:MTL}.

\subsection{Enhanced Graph Convolutional Network for Surrogate Model Construction}\label{sec:surrogate}
\begin{figure}[htbp]
    \setlength{\abovecaptionskip}{-0.1cm}  
    \setlength{\belowcaptionskip}{-0.1cm}   
    \centering
    \includegraphics[width=1\columnwidth]{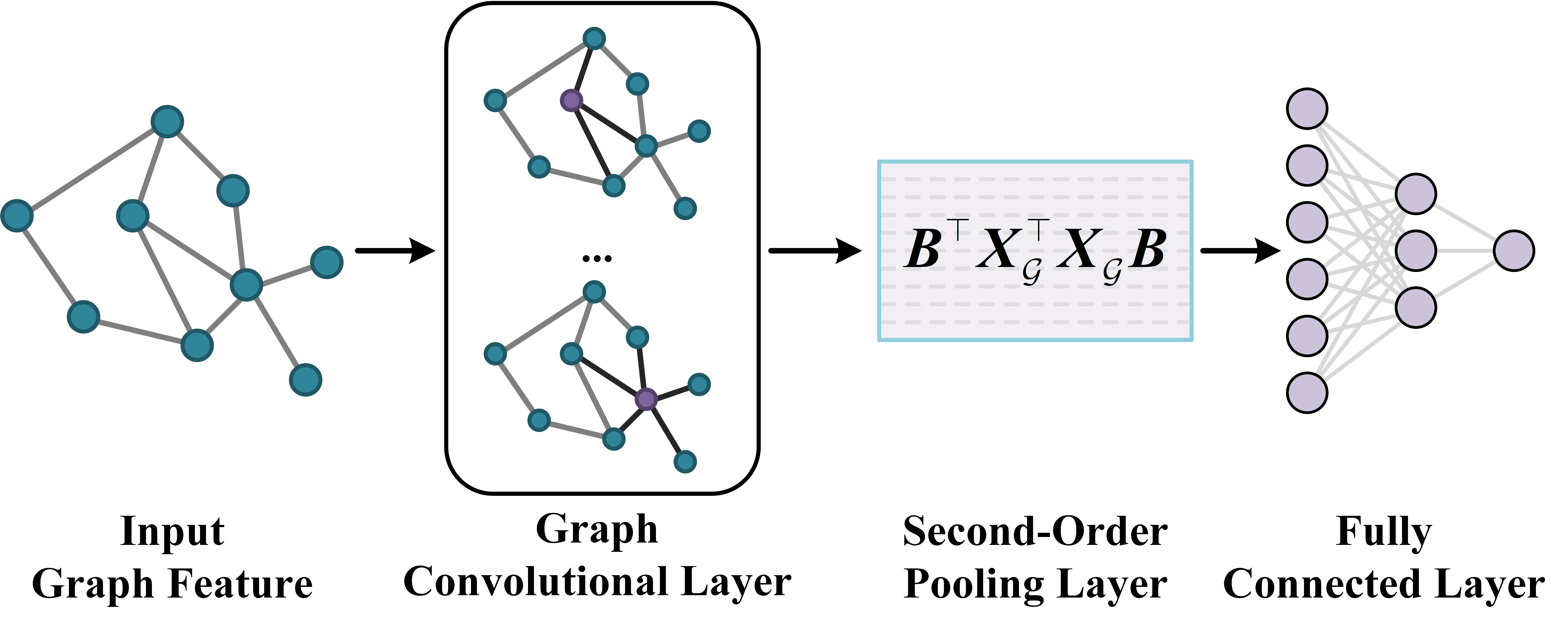}
    \caption{Structure of surrogate model constructed by enhanced GCN.}
    \label{fig:GCN}
\end{figure}

Noticing that the multiple uncertainties are spatially distributed and coupled in the power system,
the associated weights $\omega$ are inherently graph-structured due to the power network topology. 
\colorblack{
    Graph convolutional networks offer distinct advantages 
    by aligning naturally with grid topology, 
    embedding spatial correlations and node–edge dependencies through graph convolutions. 
}
Accordingly, we propose an enhanced graph convolutional network as the surrogate model to 
capture the graph coupling relationships among weights, as illustrated in Fig.~\ref{fig:GCN}.

First, the weights are organized into graph-structured data based on the power system topology.
Specifically, 
vertices correspond to buses, edges represent branches, 
the admittance matrix serves as the weighted adjacency matrix $\bm{W}$,
and the weights $\omega$ are assigned as vertex features (with zero padding for nodes without uncertain variables).
Then, the graph-structured data are fed into a spectral graph convolutional layer for feature extraction.
By performing convolutions in the Fourier domain, 
spectral GCNs~\cite{spectralGCN} effectively capture global graph information while maintaining computational efficiency.
The core operation of spectral GCN is as follows:
\begin{equation}\label{eq:GCN}
    \bm{X}\ast_{\mathcal{G}}\bm{g}_{\vartheta} := \sum_{{k}=1}^{K^\mathrm{s}} \vartheta_{k} T_{k}(\tilde{\bm{L}}) \bm{X},
\end{equation}
where $\tilde{\bm{L}} = 2\bm{L} / \lambda^{\mathrm{max}} - \bm{I}$ is the normalized adaptive graph Laplacian matrix.
$\bm{L} = \bm{I} - \bm{(D)^{-\frac{1}{2}}} \bm{W} \bm{(D)^{-\frac{1}{2}}}$.
$\bm{D}= \{D_{ii} = \sum_j W_{ij}\}$ is the degree matrix of the graph,
and $\bm{I}$ is the identity matrix.
$\lambda^{\mathrm{max}}$ is the largest eigenvalue of $\bm{L}$.
$\bm{X}$ is the input graph feature matrix.
$g_{\vartheta}$ is the filter parameterized by $\vartheta$.
$\vartheta_k$ and $T_k(\cdot)$ are the Chebyshev coefficients and polynomials of order $k$, respectively.
$K^\mathrm{s}$ is the number of Chebyshev polynomials.

The spectral GCN layer produces feature matrix $\bm{X}_{\mathcal{G}}$ for all nodes in the graph.
$\bm{X}_{\mathcal{G}}$ is then fed into a bilinear mapping second-order pooling layer~\cite{SOP} 
to further enhance the graph representation.
\begin{equation}\label{eq:pooling}
    h_{\mathcal{G}} = \text{flatten}(\bm{B}^\top\bm{X}_{\mathcal{G}}^\top\bm{X}_{\mathcal{G}}\bm{B}),
\end{equation}
where $\text{flatten}(\cdot)$ reshapes the matrix into a vector,
and $\bm{B}$ is the linear mapping matrix. 
Graph pooling aggregates node features to generate a unified graph representation. 
Unlike traditional first-order graph pooling methods (e.g., max, average, or sum pooling),
the second-order pooling layer captures second-order feature correlations and topological information across all nodes,
leading to improved representation capability,
while simultaneously reducing output dimensionality.

Finally, the graph representation vector $h_{\mathcal{G}}$ is passed through a fully connected layer 
to generate the final mapping result $\phi_{\vartheta } (\omega)$.
\colorblack{The generalization capability of the surrogate model is guaranteed by establishing a universal weight-to-PDPL mapping through the integration of power-system graph structures and training on diverse datasets.}

\subsection{Multi-Task Learning for Efficient Dataset Sample Generation}\label{sec:MTL}

\begin{figure}[htbp]
    \setlength{\abovecaptionskip}{-0.1cm}  
    \setlength{\belowcaptionskip}{-0.1cm}   
    \centering
    \includegraphics[width=0.95\columnwidth]{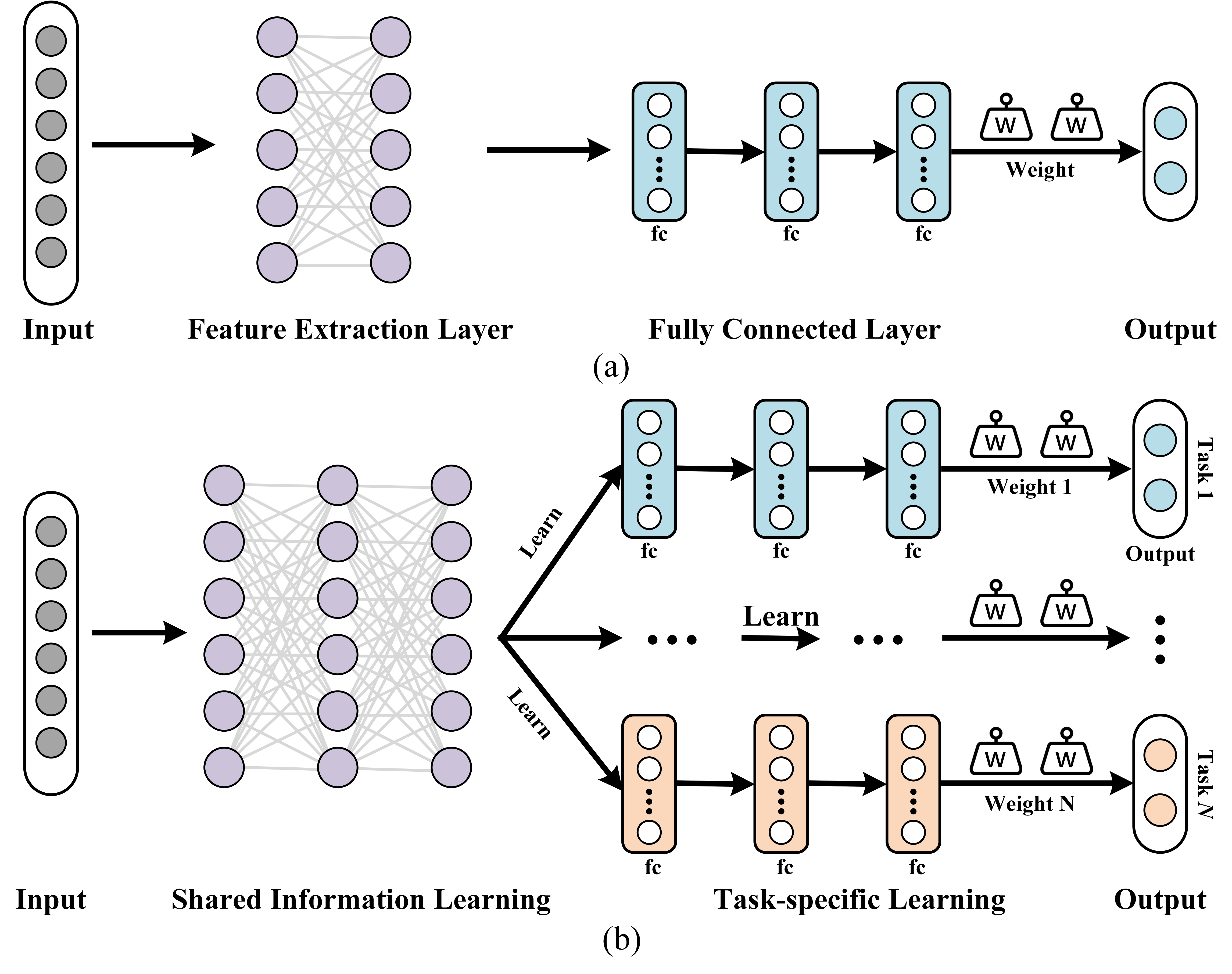}
    \caption{Model structure: (a) single-task learning and (b) multi-task learning.}
    \label{fig:MTL}
\end{figure}

\colorblack{
Each dataset sample (i.e., $(\omega, \mathcal{L}^{\mathrm{D}}_{\varphi_{\theta_{\omega}}})$) in $\mathcal{D}^{\mathrm{S}}$ 
requires training a predictive model $\varphi_{\theta_{\omega}}(\cdot)$ under the weight setting $\omega$.
In single-task learning (STL), each predictive model is trained independently, as illustrated in Fig.~\ref{fig:MTL}(a).
The loss function of STL under weight setting $\omega$ is defined as:
\begin{equation}\label{eq:STLoss}
    \mathcal{L}_{\omega}^{\mathrm{STL}} := \mathbb{E}_{(s, \xi) \in \mathcal{D}^{\mathrm{F}}}
    \big[\mathcal{L}_{\omega}(\varphi_{\theta_{\omega}}(s), \xi)\big],\  \forall \omega \in \mathcal{W}.
\end{equation} 
}
\colorblack{
However, constructing a large-scale dataset $\mathcal{D}^{\mathrm{S}}$ for surrogate model training
requires training $|\mathcal{D}^{\mathrm{S}}|$ predictive models under different weight settings,
which is computationally expensive.
}

Note that the tasks of training various predictive models with different weights exhibit significant structural similarities, 
as the primary distinction lies in the weight settings in the loss function,
\colorblack{while the structure of the predictive model and the training dataset remain unchanged.}
To leverage these similarities, 
we adopt a multi-task learning method to enable joint learning of multiple predictive models with different weights simultaneously,
\colorblack{
which significantly reduces the computational burden and training time.
}

The core idea of MTL is an information-sharing mechanism~\cite{MTL_EV}, as illustrated in Fig.~\ref{fig:MTL}(b). 
Specifically, 
a shared information learning layer serves for extracting shared features across tasks, 
thereby reducing model redundancy. 
Concurrently, independent output layers are maintained for each task to capture task-specific variations. 
This dual mechanism ensures that the shared knowledge is leveraged without compromising the unique characteristics of each task.
Note that the network structure of information-sharing layer and task-specific layer can be designed according to the form if a specific task.
Besides, the shared information learning layer in MTL can be more complex (e.g., deeper and wider) than the
feature extraction layer in STL to simultaneously handle more tasks.

Compared with traditional STL,
the information-sharing mechanism of MTL significantly reduces the number of parameters,
thereby reducing computational burden and improving training efficiency.
Suppose the parameter count of the feature extraction layer in STL is $|\theta^{\mathrm{S}}|$,
the parameter count of the shared feature extraction layer in MTL is $|\tilde{\theta}^{\mathrm{S}}|$,
and the task-specific output layer in MTL is $|\theta^{\mathrm{TS}}|$.
When $|\mathcal{W}|$ predictive models are trained simultaneously,
the parameter count of the MTL model is  $|\tilde{\theta} ^{\mathrm{S}}|+|\mathcal{W}|\times|\theta^{\mathrm{TS}}|$,
while the parameter count of the STL model is $|\mathcal{W}|\times(|\theta^{\mathrm{S}}|+|\theta^{\mathrm{TS}}|)$.

The loss function for multi-task learning is defined as:
\begin{equation}\label{eq:MTLoss}
    \mathcal{L}^{\mathrm{MTL}}_{\mathcal{W}} := 
    \frac{1}{|\mathcal{W}|}\sum_{\omega\in \mathcal{W} }  \mathbb{E}_{(s, \xi) \in \mathcal{D}^{\mathrm{F}}}
    \big[\mathcal{L}_{\omega}(\varphi_{\theta_{\omega}}(s), \xi)\big].
\end{equation}

The effectiveness of MTL can also be evaluated by the distinction of prediction performance between MTL and STL.
\begin{equation}\label{eq:MTLReduction}
    \Delta \mathcal{L}^{\mathrm{MTL}}_{\mathcal{W}} = 
        \frac{1}{|\mathcal{W}|}\sum_{\omega\in \mathcal{W} } |\mathcal{L}^{\mathrm{MTL}}_{\omega} - 
        \mathcal{L}_{\omega}^{\mathrm{STL}}|.
\end{equation} 
where $\mathcal{L}^{\mathrm{MTL}}_{\mathcal{\omega}}$ denotes the loss of the $\omega$ task in MTL.

Leveraging the proposed MTL method for large-scale predictive model training,
and through efficient optimization processes in~\eqref{eq:optimization},~\eqref{eq:decisionLoss} and~\eqref{eq:expectedDecisionLoss},
$\mathcal{D}^{\text{S}} $ can be efficiently generated.

\subsection{Algorithm of WPO Framework}\label{sec:algorithm}
The algorithm of the proposed WPO framework is summarized in Algorithm~\ref{algorithm1}.

\begin{algorithm}[htbp]\label{algorithm1}
    \caption{Weighted Predict-and-Optimize}
    \SetAlgoLined
    \SetEndCharOfAlgoLine{}
    \KwIn{Prediction dataset $\mathcal{D}^{\mathrm{F}}$ of uncertainty realizations \\$\xi$ and corresponding features $s$.}  
    \KwOut{Optimized weights $\omega^*$ for critical uncertainties.}  
    
    \SetKwBlock{StepOne}{Step 1 - Surrogate Dataset $\mathcal{D}^{\text{S}}$ Construction}{}
    \SetKwBlock{StepTwo}{Step 2 - Surrogate Model Training}{}
    \SetKwBlock{StepThree}{Step 3 - Weight Optimization}{}
    
    \StepOne{  
        Generate a set of weight settings $\mathcal{W}$.  
    
        Train the predictive models using the MTL method.
    
        Accumulate data pairs ($\omega$, $\mathcal{L}^{\mathrm{D}}_{\varphi_{\theta_{\omega}}}$) by:
        
        \For{$\omega\in \mathcal{W}$}{
            Predict all $\hat{\xi}=\varphi_{\theta_{\omega}}(s)$ in $\mathcal{D}^{\mathrm{F}}$.
    
            Compute $\mathcal{L}^{\mathrm{D}}_{\varphi_{\theta_{\omega}}}$ 
            through \eqref{eq:optimization}-\eqref{eq:expectedDecisionLoss}.
     }  
    }  
    
    \StepTwo{  
        Construct the surrogate model using enhanced GCN.

        Train the surrogate model on $\mathcal{D}^{\text{S}}$ to obtain $\phi_{\vartheta^\ast}(\cdot)$.
    }  
    
    \StepThree{  
        Optimize the weights of critical uncertainties by~\eqref{eq:weightUpdate}
        \\on $\phi_{\vartheta^\ast}(\cdot)$ and obtain $\omega^*$.
    }  
    
\end{algorithm}

It's important to note that 
the proposed WPO framework is applicable to both deterministic prediction-deterministic optimization and 
probabilistic prediction-stochastic optimization problems.
For clarity and simplicity, we focus on deterministic prediction–deterministic optimization in the following case study.
Accordingly, we adopt $\ell(\hat{\xi}, \xi) = (\hat{\xi} - \xi)^2$. 
In the next section, we present the detailed formulation of a classical deterministic optimization problem in the distribution network.

\section{Case Study on Optimal Operation \\in Distribution Network}\label{sec:risk}

In this section, we consider a classic predict-optimize problem:  
optimal dispatch of distributed generation (DG)  in a distribution network (DN).
In the DN, certain nodes are integrated with DGs and uncertain loads (ULs) (e.g., electric vehicle charging stations),
while the load demands at other nodes are assumed fixed for simplicity.
The uncertain loads introduce potential risks to the DN operation.
In this paper, we focus on the voltage drop below the limit as the primary risk.
Leveraging the voltage support capability of DGs,
DN aims to optimize the DG dispatch strategy to manage the potential risks. 
First, the uncertain load demands are predicted for uncertainty quantification.
Then, based on the predictions,
DN optimizes the DG dispatch strategy to minimize operational costs while ensuring secure network operation.
After the true realizations of the uncertain loads are revealed,
DN adopts additional resources to manage the risks caused by prediction deviations
and thus facing additional economic costs.

\subsection{Objective}
The objective is to minimize the total operation cost including the DG operation cost and the trading cost with the transmission network.
\begin{equation}
        \min\  C^{\mathrm{o}}=\pi^{\mathrm{T}}P^{\mathrm{T}}+\pi^{\mathrm{G}} \sum\limits_{i\in \varOmega^{\mathrm{G}}}P_{i}^{\mathrm{G}}\\
\end{equation}
where $\varOmega^{\mathrm{G}}$ refers to the set of nodes with DGs.
$\pi^{\mathrm{G}}/\pi^{\mathrm{T}}$ are the cost coefficients of DG operation and power trading, respectively.
$P_{i}^{\mathrm{G}}$ is the output of DG at node $i$.
$P^{\mathrm{T}}$ is the trading power.

\subsection{Constraints}

\subsubsection{Power Flow Constraints}
The distflow model is used to describe the power flow constraints in the DN.
We denote $\varOmega^{\mathrm{N}}/\varOmega^{\mathrm{B}}$ as the set of nodes/branches.
For $\forall i \in \varOmega^{\mathrm{N}}, \forall ij \in \varOmega^{\mathrm{B}}$, we have:
\begin{subequations}\label{pf}
  \begin{align}
      \label{pf U}
      & V_{j}^2 = V_{i}^2 - 2(r_{ij}P_{ij} + x_{ij}Q_{ij})+(r_{ij}^2 + x_{ij}^2)I_{ij}^2 \\
      \label{pf P}
      & p_{j} = P_{ij}-r_{ij}I_{ij}^2-\sum_{l:j\rightarrow l}P_{jl} \\
      \label{pf Q}
      & q_{j} = Q_{ij}-x_{ij}I_{ij}^2-\sum_{l:j\rightarrow l}Q_{jl} \\
      \label{pf nonconvex}
      & V_{i}^2I_{ij}^2 = (P_{ij})^2+(Q_{ij})^2 \\
      \label{pf U range}
      & \underline{V}\leq V_{i}\leq \overline{V} \\
      \label{pf I range}
      & |I_{ij}| \leq\overline{I}_{ij} 
  \end{align}
\end{subequations}
where $r_{ij}/x_{ij}$ are the line resistance/reactance of line $ij$, respectively.
$I_{ij}$ is the electric current of line $ij$, 
with $\overline{I}_{ij}$ as the upper line current.
$V_{i}$ is the voltage of bus $i$, with $\overline{V}/\underline{V}$ as the upper/lower bus voltage bounds.
$P_{ij}/Q_{ij}$ are the line active/reactive power of line $ij$, respectively.
$p_{i}/q_{i}$ are the active/reactive outflow power of bus $i$.
\eqref{pf U} describes the voltage drop over line $ij$. 
\eqref{pf P} and~\eqref{pf Q} represent the active and reactive power balance of bus $j$.
\eqref{pf nonconvex} is the power flow equation of line $ij$.
\eqref{pf U range},~\eqref{pf I range} are the security constraints.

Notably, the non-convex constraint~\eqref{pf nonconvex} can be relaxed to the following second-order cone formulation 
by introducing two auxiliary variables $\mathsf{V}_{i}$ and $\mathsf{I}_{ij}$ to replace the quadratic term, 
as in~\eqref{SOCP}. 
\begin{subequations}\label{SOCP}
    \begin{align}
        &\left \| \begin{matrix}2P_{ij}
            \\2Q_{ij}
            \\\mathsf{V}_{i} - \mathsf{I}_{ij}
        \end{matrix} \right \| _2 \leq \mathsf{V}_{i} + \mathsf{I}_{ij}\\
        &\mathsf{V}_{i} = V_{i}^2,\ \mathsf{I}_{ij} = I_{ij}^2
    \end{align}
\end{subequations}

\subsubsection{Energy Balancing Constraints}
For $\forall i \in \varOmega^{\mathrm{N}}$, we have
\begin{subequations}\label{energybalance}
    \begin{align}
        \label{pin}
        & p_{i} = P_{i}^{\mathrm{L}}  + \hat{P}_{i}^{\mathrm{UL}} - P_{i}^{\mathrm{G}} \\
        \label{qin}
        & q_{i} = Q_{i}^{\mathrm{L}}
    \end{align}
\end{subequations}
where 
$P_{i}^{\mathrm{L}}/Q_{i}^{\mathrm{L}}$ are the active/reactive load power of bus $i$.
$\hat{P}_{i}^{\mathrm{UL}}$ is the predicted uncertain loads at node $i$.
We denote $\varOmega^{\mathrm{UL}}$ as the set of nodes with uncertain loads.
Notably, in~\eqref{pin}, $P_{i}^{\mathrm{G}}=0$ if $i \notin \varOmega^{\mathrm{G}}$,
and $\hat{P}_{i}^{\mathrm{UL}}=0$ if $i \notin \varOmega^{\mathrm{UL}}$.

\subsubsection{DG Operation Constraint}
For $\forall i \in \varOmega^{\mathrm{G}}$, we have
\begin{equation}\label{DG}
  0\leq P_{i}^{\mathrm{G}} \leq \overline{P_{i}}^{\mathrm{G}}
\end{equation}
where $\overline{P_{i}}^{\mathrm{G}}$ is the upper limit of DG output at node $i$.

\subsection{Overall Problem}
Under the predictions of the uncertain loads, the optimization problem can be formulated as follows:
\begin{equation}\label{overall prob1}
    \begin{split}
        \min_{\Xi}\  &C^{\mathrm{o}}=\pi^{\mathrm{T}}P^{\mathrm{T}}+\pi^{\mathrm{G}} \sum\limits_{i\in \varOmega^{\mathrm{G}}}P_{i}^{\mathrm{G}}\\
        \text{s.t.}\  & \eqref{pf U}-\eqref{pf Q},\ \eqref{pf U range}-\eqref{pf I range},\ 
        \eqref{SOCP}-\eqref{DG}
    \end{split}
\end{equation}
where $\Xi=\{P_{i}^{\mathrm{G}},P^{\mathrm{T}}|i\in \varOmega^{\mathrm{G}}\}$ is the decision variable set.

DN first makes operation decisions $\Xi$ based on the predictions of uncertain loads through~\eqref{overall prob1}.
Upon the revelation of the true values of uncertain loads, 
$\Xi$ is implemented in the system. 
However, due to unavoidable prediction deviations, 
$\Xi$ may not be fully compatible with true operating environment, 
potentially exposing the DN to additional operational risks. 
To manage these risks, 
supplementary resources (e.g., reserves, load shedding) must be deployed, 
leading to increased operational costs, which can be simply computed as
$C^{\mathrm{p}} = \pi^{\mathrm{p}}\sum_{i\in \varOmega^{\mathrm{N}}}[\underline{V}-V_i]^+$.
$[\cdot]^+$ denotes the positive part of the argument.
$\pi^{\mathrm{p}}$ is the penalty coefficient for voltage violations. 
The overall cost is $C= C^{\mathrm{o}} + C^{\mathrm{p}}$.

\section{Numerical Case Study}\label{sec:case}

\subsection{Set Up}
\begin{figure}[htbp]
    \setlength{\abovecaptionskip}{-0.1cm}  
    \setlength{\belowcaptionskip}{-0.1cm}   
    \centering
    \includegraphics[width=1\columnwidth]{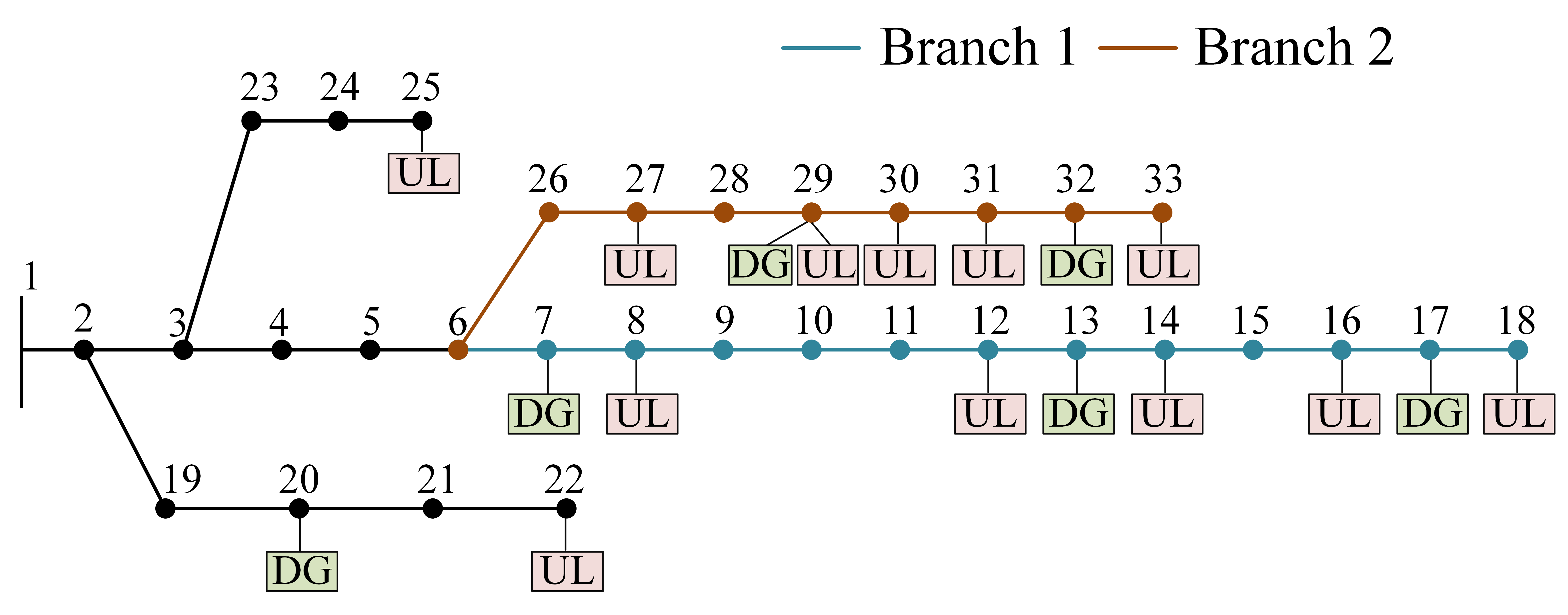}
    \caption{Modified IEEE 33-bus DN with DGs and uncertain loads.}
    \label{fig:IEEE33}
\end{figure}

A modified IEEE 33-bus distribution network integrated with DGs and uncertain loads
is used as the test system, as illustrated in Fig.~\ref{fig:IEEE33}.
$\varOmega^{\mathrm{UL}}=$\{8, 12, 14, 16, 18, 22, 25, 27, 29, 30, 31, 33\}.
$\varOmega^{\mathrm{G}}=$\{7, 13, 17, 20, 29, 32\}.
The voltage magnitude is constrained as $|V_i| \in [0.90,1.10]\ \text{(p.u.)},\ \forall i \in \varOmega^\mathrm{N}$. 
To emphasize potential low voltage challenges, the original nodal loads are scaled up by a factor of 1.05.
$\pi^{\mathrm{G}}=\$10 $/MWh, $\pi^{\mathrm{T}}=\$20 $/MWh, 
$\pi^{\mathrm{p}}=\$100 $/MWh. 
We set $\overline{P_{i}}^{\mathrm{G}}=1\ \text{(p.u.)},\ \forall i \in \varOmega^\mathrm{G}$ to ensure 
the DGs can provide adequate voltage support.
$\mathcal{W}$ is generated by sampling from a Dirichlet distribution.
\colorblack{
In this study, we adopt the spatio-temporal graph convolutional network (STGCN)~\cite{PTO_EVpredict}
as the predictive model to capture the spatio-temporal correlations of uncertain loads.
Thus, the information sharing layer in MTL is designed as STGCN,
and the task-specific layer is designed as a multilayer perceptron (MLP).
The detailed parameters of MTL are provided in~\cite{data}.
The information set $s$ includes historical load demand observations,
the prediction horizon is set as 1 time step (i.e., 1 quarter).
The parameters of the enhanced GCN model are set as: 
Number of nodes = 33,
$K^{\mathrm{s}} = 3$,
learning rate = $0.003$,
fc-blocks = $[8\times8, 16, 1]$.
}
The deep learning models are implemented in Python 3.10.11 using PyTorch 2.5.0 with CUDA 12.6 support.
The optimization models are implemented in Python with the Gurobipy interface and solved by Gurobi 12.0 solver. 
All computations are conducted on a Windows 11 64-bit operating system equipped with an Intel Core i9-13900HX @ 2.30 GHz processor, and 16 GB RAM.

\subsection{Performance of Weighted Prediction}
\begin{figure}[htbp]
    \setlength{\abovecaptionskip}{-0.1cm}  
    \setlength{\belowcaptionskip}{-0.1cm}   
    \centering
    \includegraphics[width=1\columnwidth]{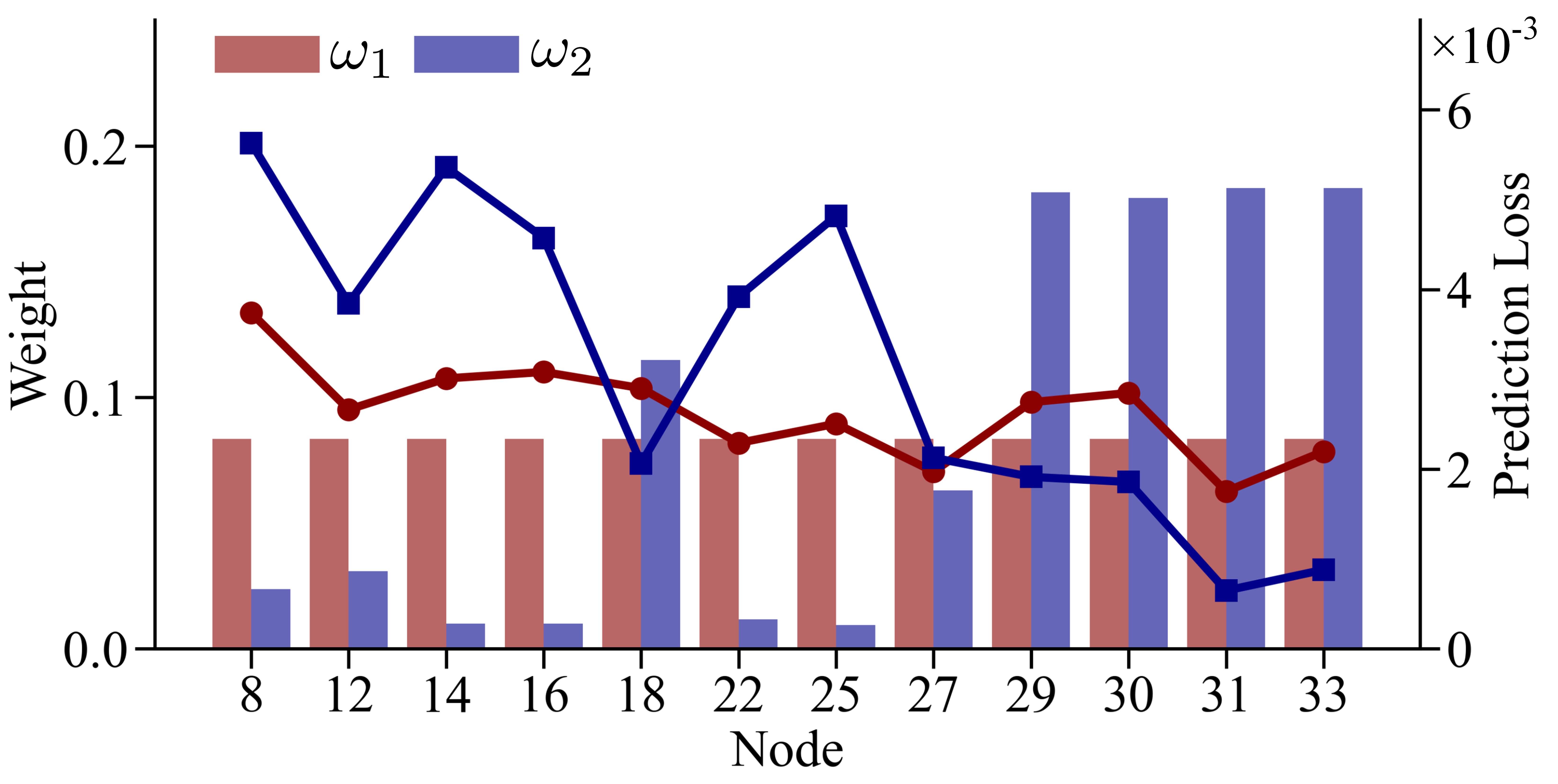}
    \caption{Node-wise prediction loss under two weight settings.}
    \label{fig:assumption1}
\end{figure}

WPO is built on the fundamental assumption that the employed predictive model exhibits satisfactory performance.
We verify this by comparing the node-wise prediction loss measured 
by $\mathbb{E}_{\mathcal{D}^\mathrm{F}}[(\hat{\xi}_i - \xi_i)^2]$ under two distinct weight settings. 
Specifically,
$\omega_1$ is uniformly assigned as $1/|\varOmega^{\mathrm{UL}}|$, corresponding to the traditional predict-then-optimize paradigm.
$\omega_2$ is generated by sampling from a Dirichlet distribution, representing a weighted prediction.
Fig.~\ref{fig:assumption1} shows the relationship between the nodal weights (bar chart) 
and the corresponding node-wise prediction losses (line chart).

Firstly, the prediction losses of $\omega_1$ and $\omega_2$ are both within an acceptable range,
verifying that the weighted predicting method does not compromise prediction accuracy.
Secondly, compared with $\omega_1$,
the node-wise prediction losses of $\omega_2$ tend to be smaller with higher weights (nodes 18, 29, 30, 31, 33).
\colorblack{This phenomenon is attributed to the fact that, 
during the gradient descent training, 
prediction variables with larger weights contribute more to the overall loss gradient.
Consequently, 
the training process inherently prioritizes reducing the loss of high-weight variables, 
thereby improving their relative prediction accuracy.
}

\subsection{Performance of WPO Framework}\label{secV:WPO}
In this subsection, we first evaluate the performance of WPO framework in two test cases.
Then, we analyze the performance of the enhanced GCN in achieving accurate surrogate-model mapping and evaluate the MTL’s effectiveness in large-scale predictive model training.

\subsubsection{Performance of Weight Optimization}
Leveraging the trained surrogate model, 
we optimize the weight settings of critical uncertainties to minimize the PDPL
by performing gradient descent on the surrogate model as in~\eqref{eq:weightUpdate}.
Two test cases are designed to represent distinct risk scenarios under different integration levels of uncertain loads:

\begin{enumerate}
    \item \textbf{Case 1}: The integration levels of uncertain loads in $\varOmega^{\mathrm{UL}}$ are relatively uniform. 
    Based on the original power flow characteristics of the 33-bus network, 
    voltage drop risks primarily concentrate on the branch associated with node 18 (branch 1 in Fig.~\ref{fig:IEEE33}).
    \item \textbf{Case 2}: The integration levels of uncertain loads on branch 1 are relatively low, 
    while the integration levels of uncertain loads on the branch containing node 33 (branch 2 in Fig.~\ref{fig:IEEE33}) are relatively high. 
    Consequently, the voltage drop risk is relatively higher on branch 2, whereas it is lower on branch 1.
\end{enumerate}

\begin{figure}[htbp]
    \setlength{\abovecaptionskip}{-0.1cm}  
    \setlength{\belowcaptionskip}{-0.1cm}   
    \centering
    \includegraphics[width=0.95\columnwidth]{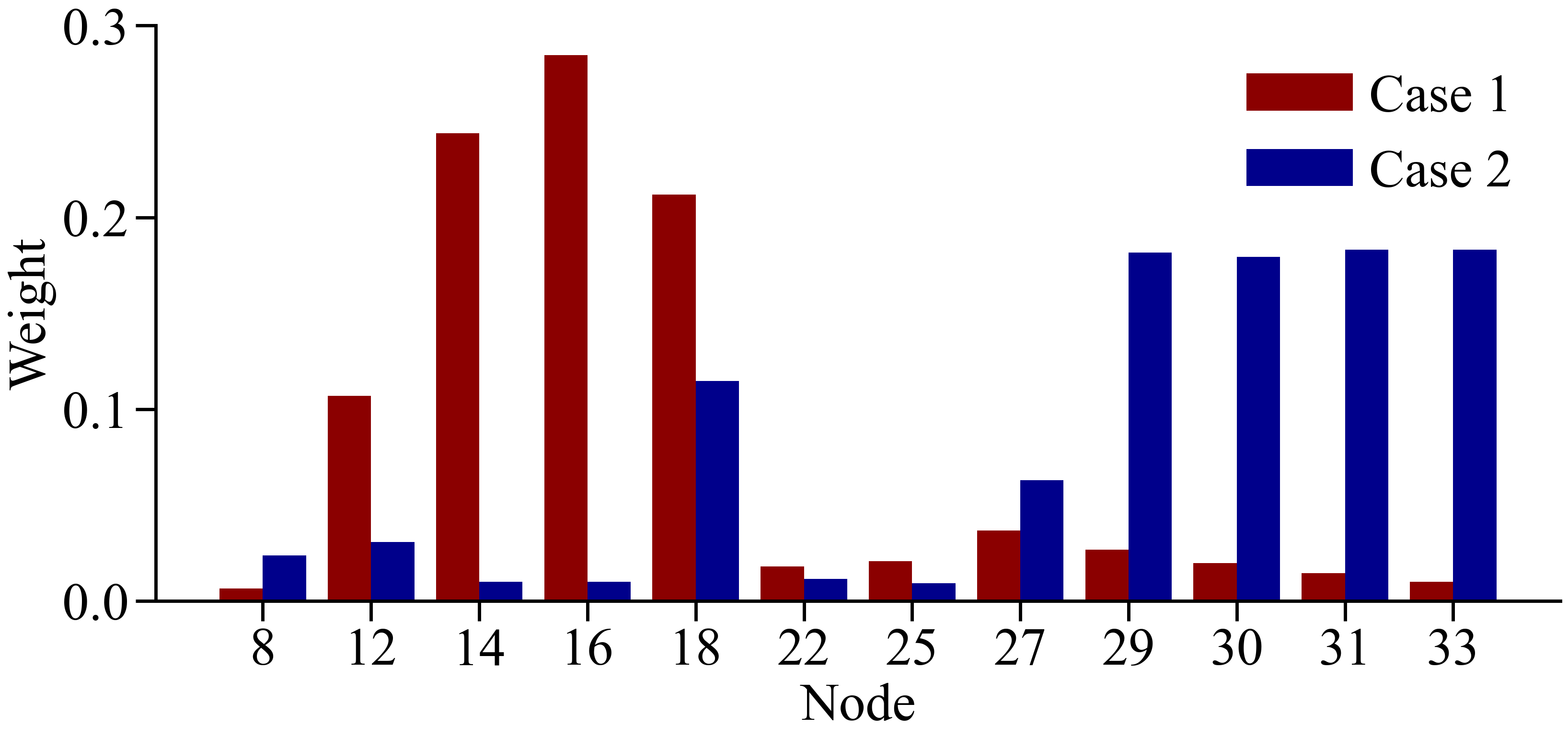}
    \caption{Results of WPO framework: red bars - Case 1, blue bars - Case 2.}
    \label{fig:weight}
\end{figure}

For each case, the WPO framework is applied to optimize the weight settings of critical uncertainties.
The optimized weight settings are shown in Fig.~\ref{fig:weight}.

Fig.~\ref{fig:weight} illustrates that the weight optimization results obtained from the WPO framework
align well with the risk profiles in each case. 
In Case 1, 
where voltage risks are primarily concentrated on branch 1, 
nodes along this branch are assigned higher weights. 
This allocation underscores the necessity of accurate predictions at these critical nodes to ensure reliable system operation.
Conversely, nodes with inherently high voltage secure margins, such as nodes 8, 22, and 25, receive significantly lower weights, 
reflecting small impacts of their prediction deviations on the overall optimization outcomes. 
Similarly, nodes in branch 2, where no significant voltage risk is observed in this case, 
are also assigned lower weights.
Interestingly, despite the elevated risk level at node 18, it is not assigned the highest weight. 
Instead, nodes 14 and 16 receive higher weight allocations. 
This is attributed to their positioning within the network: located at the end of the feeder and upstream of node 18,
where prediction deviations can propagate and exacerbate voltage risks. 
As inaccuracies at nodes 14 and 16 directly affect the voltage profile of node 18, 
the WPO framework prioritizes these nodes with higher weights to enhance overall predictive robustness.

In Case 2, 
the risk level along branch 2 surpasses that of branch 1. 
This shift is primarily attributed to load redistribution: an increase in load along branch 2 intensifies its voltage risk, 
whereas a reduction in load on branch 1 alleviates its associated risks. 
Consequently, only the terminal node 18 experiences a slight voltage drop below the acceptable threshold, 
while the overall risk along branch 1 significantly reduced.
This redistribution of voltage risks is directly reflected in the weight optimization results. 
Nodes along branch 2 are assigned higher weights, 
emphasizing their critical role in ensuring accurate predictions and maintaining system safety. 
In contrast, 
nodes along branch 1 receive lower weights, 
as their reduced load levels mitigate voltage risks. 
However, 
node 18 retains a relatively higher weight than other nodes in branch 1 due to its pivotal position within the network.
Overall, 
these findings validate the effectiveness of the proposed WPO framework in adaptively aligning weight allocations with system risk profiles.

\subsubsection{Performance of Surrogate Model}\label{secV:surrogate}
\begin{figure}[htbp]
\color{blue}
    \setlength{\abovecaptionskip}{-0.1cm}  
    \setlength{\belowcaptionskip}{-0.1cm}   
    \centering
    \includegraphics[width=0.95\columnwidth]{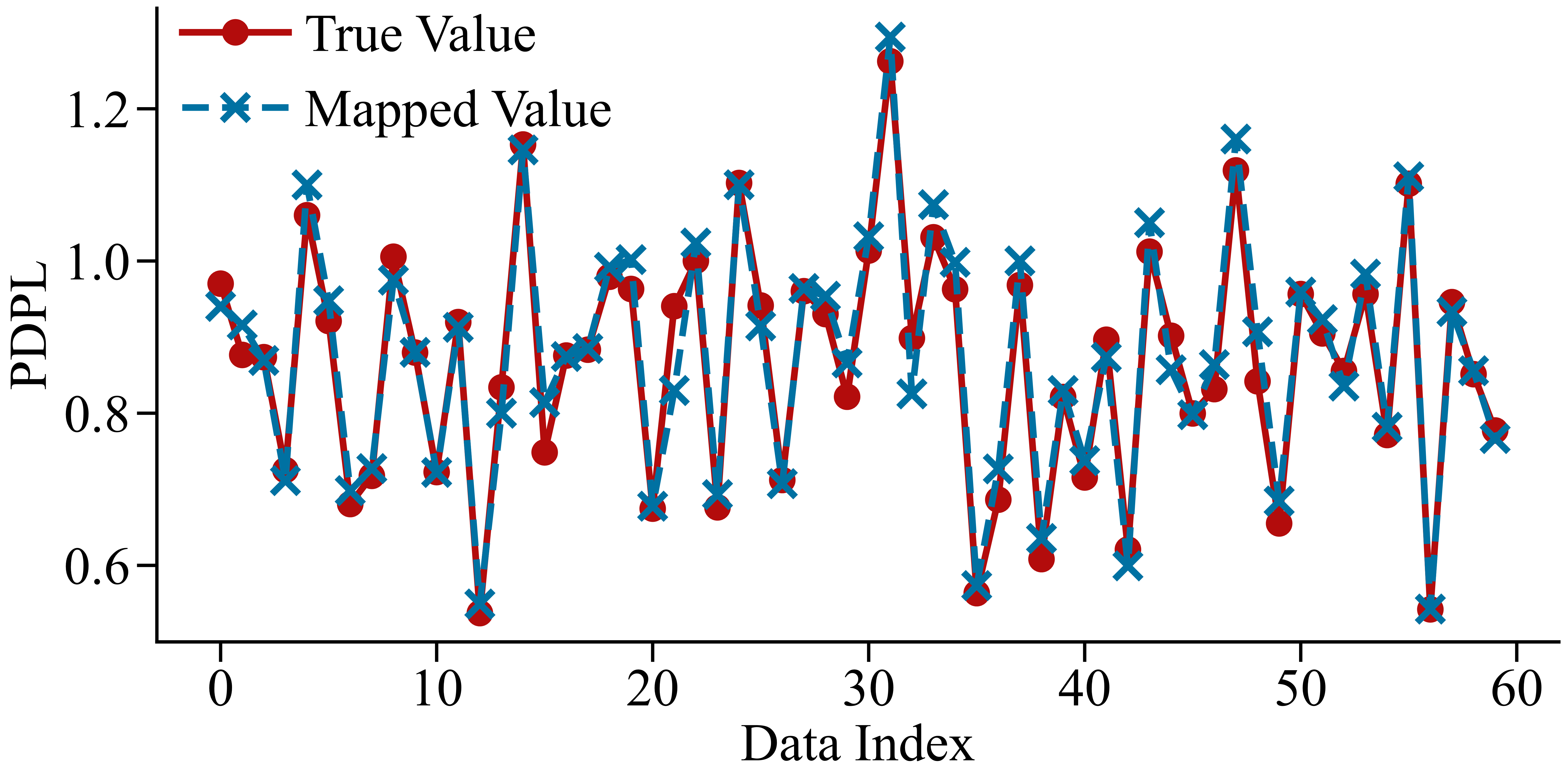}
    \caption{Mapping performance of the surrogate model.}
    \label{fig:GCNmapping}
\end{figure}

A dataset $\mathcal{D^{\text{S}}}$ containing 10,000 samples is used to train the enhanced GCN model.
\colorblack{
To illustrate the surrogate model’s mapping capability,
we visualize the mapping performance on the test dataset in Fig.~\ref{fig:GCNmapping}.
}
The surrogate model achieves a small mapping loss of $8\times 10^{-4}$ on the test dataset, 
demonstrating its high mapping accuracy. 
This small loss highlights the high precision of the surrogate model in capturing the underlying data patterns.

\begin{figure}[htbp]
    \setlength{\abovecaptionskip}{-0.1cm}  
    \setlength{\belowcaptionskip}{-0.1cm}   
    \centering
    \includegraphics[width=0.95\columnwidth]{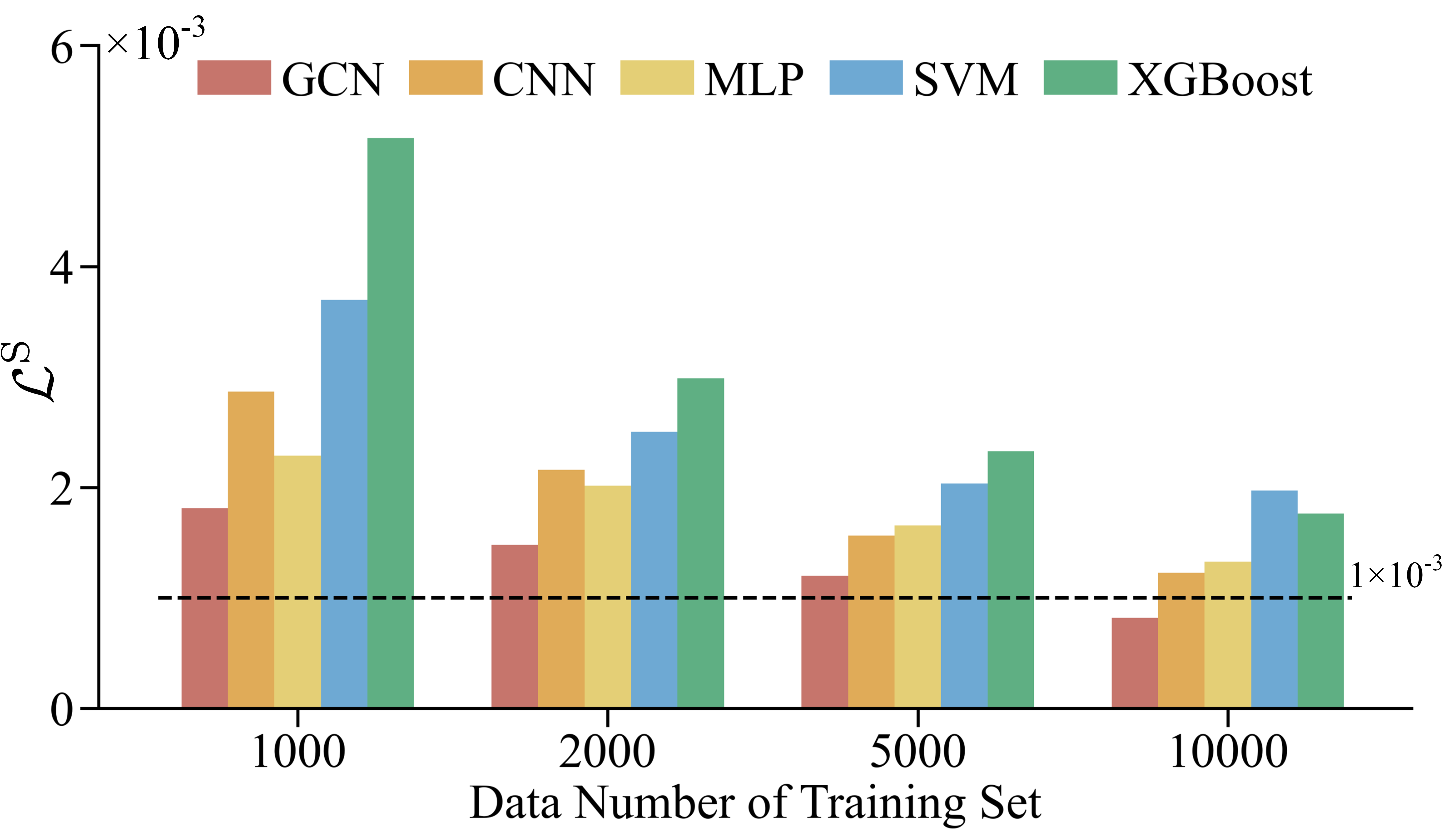}
    \caption{Comparison of mapping performance of different models.}
    \label{fig:mappingCompare}
\end{figure}

While the proposed MTL method significantly alleviates the computational burden associated with training multiple prediction tasks, 
the computation cost remains non-negligible.
Therefore, 
it is essential to determine the minimum dataset size required to train the surrogate model while maintaining a specified error tolerance. 
To this end,
we perform a sensitivity analysis on the dataset size of $\mathcal{D^{\text{S}}}$ to assess
the impacts of varying sample numbers on the surrogate model's mapping performance.
Furthermore, 
we conduct a comparative analysis against conventional machine learning methods, 
including MLP, 
convolutional neural network (CNN), 
support vector machine (SVM), 
and XGBoost. 
The results are presented in Fig.~\ref{fig:mappingCompare}.

Fig.~\ref{fig:mappingCompare} illustrates that as the size of the training dataset increases, 
the loss values of all models generally decrease, 
indicating that larger datasets enhance model performance. 
In particular, for larger datasets (e.g., 10,000 samples), 
the loss of the GCN model is below $1 \times 10^{-3}$,
demonstrating its strong ability in capturing underlying data patterns.
Notably, the proposed GCN model consistently outperforms other models, 
achieving the lowest loss across all dataset sizes. 
In contrast, traditional models generally exhibit higher loss values,
with pronounced performance degradation in small-data regimes,
indicating their strong dependence on dataset size and their limited ability to extract complex features when data availability is constrained. 
The proposed enhanced GCN model,
by aggregating neighborhood information, capturing second-order feature correlations to enrich graph representation, 
and performing multi-layer feature fusion, 
effectively capture complex dependencies and global interactions among graph-structured variables. 
Thus, the proposed GCN model demonstrates superior generalization ability and robust mapping performance.

\begin{figure}[htbp]
\color{blue}
    \setlength{\abovecaptionskip}{-0.1cm}  
    \setlength{\belowcaptionskip}{-0.1cm}   
    \centering
    \includegraphics[width=0.95\columnwidth]{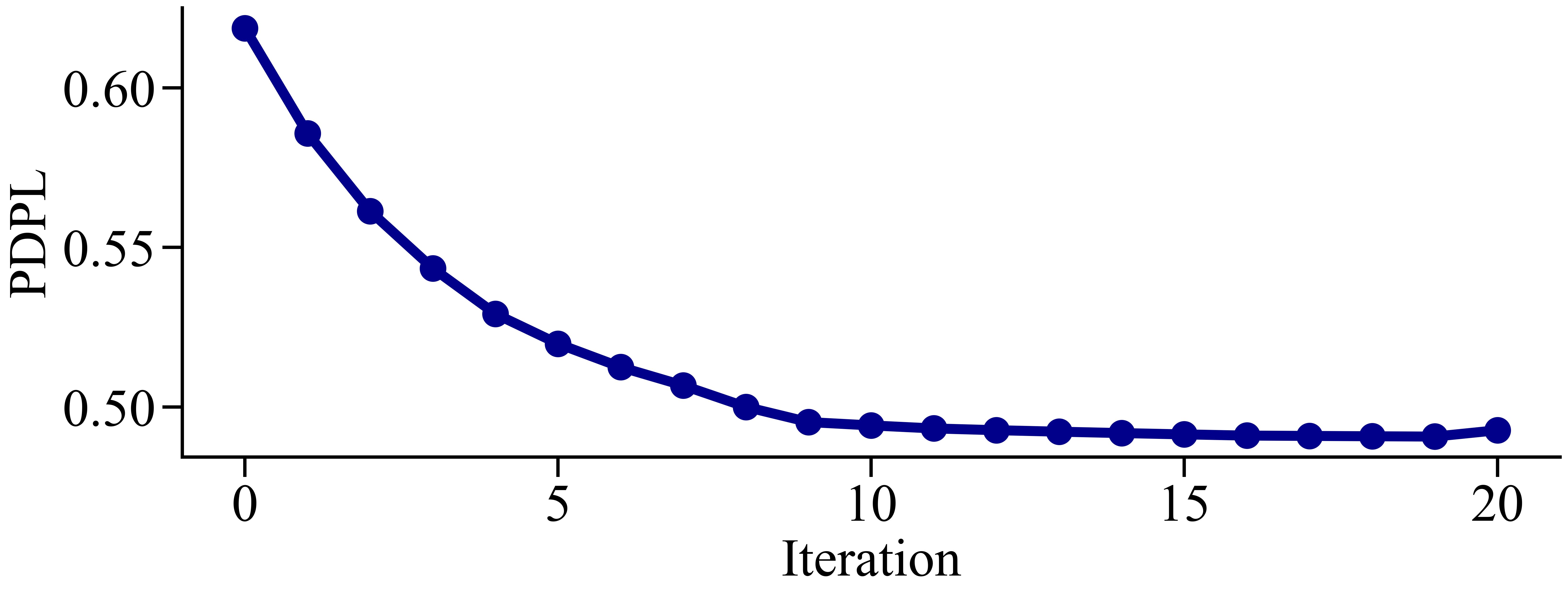}
    \caption{Convergence curve of weight optimization for Case 1.}
    \label{fig:convergence}
\end{figure}

\colorblack{
In addition to the mapping performance,
the convergence performance of~\eqref{eq:weightUpdate} is essential to guarantee that
the weight optimization process can be efficiently completed. 
The convergence trajectory of the weight optimization for Case 1 is shown in Fig.~\ref{fig:convergence}.
Fig.~\ref{fig:convergence} indicates that the weight optimization process converges rapidly to a small PDPL loss value
within 15 iterations, 
providing a solid guarantee for the efficiency of the WPO framework.
}

\subsubsection{Performance of Multi-Task Learning}
We first employ a test case of simultaneously training 100 prediction tasks with 100 weights, i.e., $|\mathcal{W}|=100$.
The evaluation metrics include 
prediction accuracy, parameter count, and training time.
A comprehensive comparison is presented in Table~\ref{tab:MTL}.

\begin{table}[htbp]
    \setlength{\abovecaptionskip}{-0.1cm}
    \setlength{\belowcaptionskip}{-0.1cm}
    \renewcommand\arraystretch{1.5} 
    \setlength{\tabcolsep}{0.18cm} 
    \caption{Comparison Between MTL and STL for $|\mathcal{W}|=100$}\label{tab:MTL}
    \begin{center}
    \begin{tabular}{cccc}
    \toprule 
                               &Prediction Accuracy      &Parameter Count       &Training Time (s) \\ 
    \midrule
    \textbf{STL}           &$2.45\times 10^{-3}$&$2,239,500$&$45,209$\\ 
    \textbf{MTL}           &$2.52\times 10^{-3}$&$296,902$&$3,571$\\ 

    \bottomrule   
    \end{tabular}
    \end{center}
\end{table}

Regarding the prediction accuracy,
both STL and MTL achieve comparable and satisfactory performance,
with losses of $2.45\times 10^{-3}$ for STL and $2.52\times 10^{-3}$ for MTL.
Furthermore, the discrepancy between MTL and STL ($\Delta \mathcal{L}^{\mathrm{MTL}} = 1.1\times 10^{-4}$)
is minimal, indicating that the integration of shared information learning in MTL does not compromise predictive performance.
In terms of the computational efficiency,
the superiority of MTL over STL is evident.
Firstly, the computation burden of MTL is significantly lower than that of STL.
For the adopted STGCN and MLP model, $|\theta^{\mathrm{S}}|=19,850$, $|\theta^{\mathrm{TS}}|=2,545$,
and $|\tilde{\theta}^{\mathrm{S}}|=42,402$.
Thus, the parameter count of MTL is 296,902, while that of STL is 2,239,500,
which is 7.54 times larger than MTL.
This is because the information-sharing mechanism inherent in MTL effectively reduces the overall parameter count, 
minimizing model computation burden while maintaining task-specific adaptability.
Moreover, MTL substantially reduces training time by eliminating redundant computations. 
MTL requires only 3,571 seconds to train a single highly integrated model, 
whereas STL requires 45,209 seconds to train 100 models, 
making MTL 12.7 times faster. 
The above results demonstrate the effectiveness of MTL in delivering comparable prediction accuracy
while significantly reducing training time and computational burden compared to STL.

\begin{figure}[htbp]
    \setlength{\abovecaptionskip}{-0.1cm}  
    \setlength{\belowcaptionskip}{-0.1cm}   
    \centering
    \includegraphics[width=0.95\columnwidth]{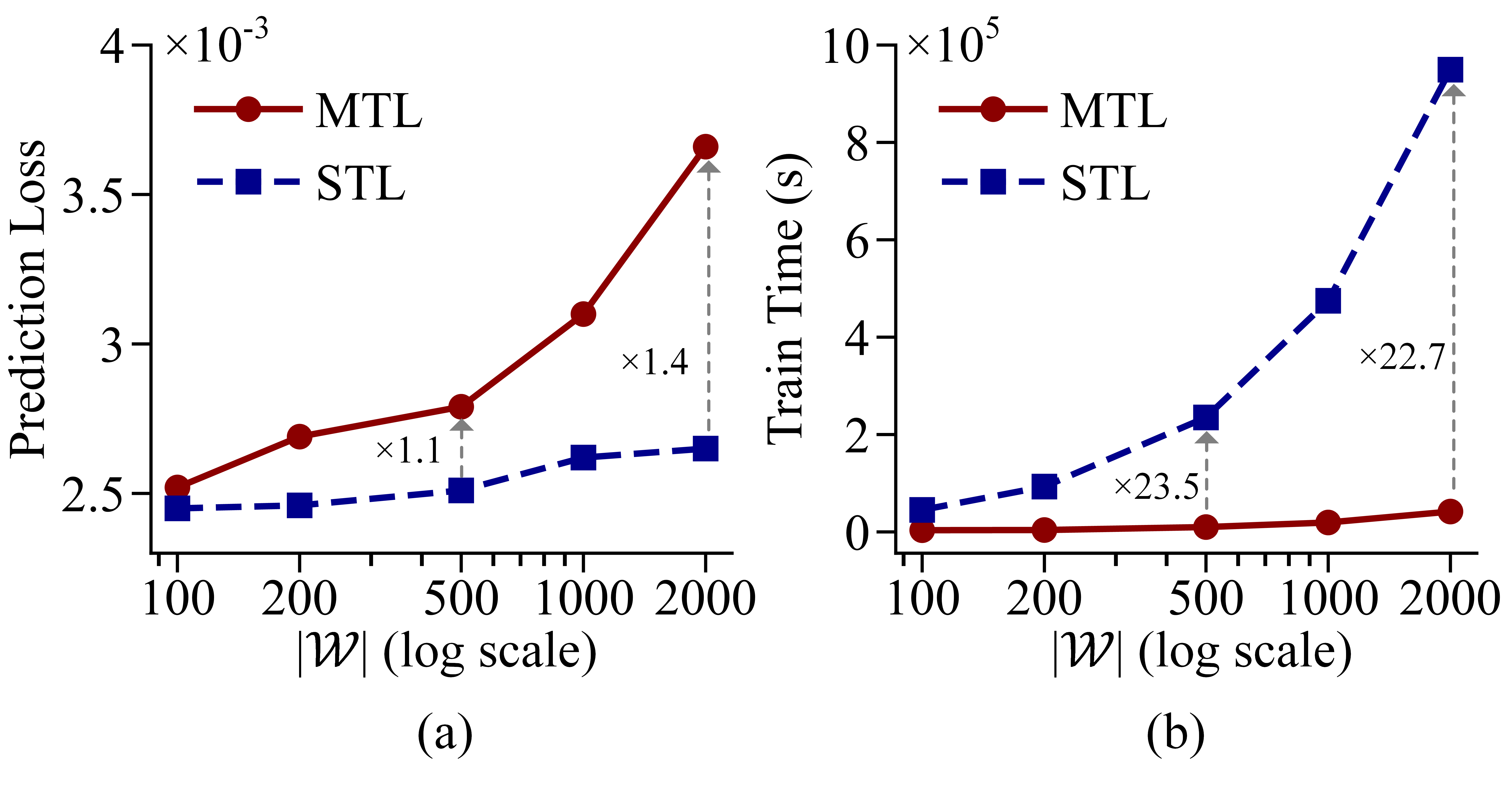}
    \caption{Scalability of MTL and STL: (a) Prediction loss, (b) Training time.}
    \label{fig:MTLSTL}
\end{figure}

To validate the scalability of MTL,
we further extend $|\mathcal{W}|$ to a larger scale,
ranging from 100 to 2,000.
The impacts of increasing $|\mathcal{W}|$ on prediction performance and training time are presented in Fig.~\ref{fig:MTLSTL}.

Fig.~\ref{fig:MTLSTL} indicates that 
as $|\mathcal{W}|$ increases,
the prediction loss of MTL gradually increases,
but remains at an acceptable range.
Although the prediction loss of STL remains in small changes as $|\mathcal{W}|$ increases,
the computational burden of STL grows exponentially.
In contrast, MTL requires substantially less training time.
When $|\mathcal{W}|=500$, MTL exhibits an 11\% increase in prediction loss relative to STL, 
yet requires only 4.25\% of STL's training time. 
As $|\mathcal{W}|$ further increases to 2,000, the prediction loss of MTL reaches 1.4 times that of STL, 
whereas its training time remains only 4.4\% of STL's.
These results demonstrate MTL's superior scalability and efficiency in handling multiple prediction tasks.

\subsection{Comparison with Alternative Weight Setting Methods}

To further evaluate the effectiveness of the proposed WPO framework,
we conduct a comparative analysis against the following weight-setting methods:

\textbf{W1:} Weights are assigned uniformly to all uncertain variables as $1/|\varOmega^{\mathrm{UL}}|$,
corresponding to the traditional predict-then-optimize paradigm.

\textbf{W2:} Weights are determined based on the voltage safety margin. 
Specifically, we first run the power flow calculation on the original IEEE 33-bus system
without the integration of DGs and uncertain loads.
For each node $i \in \varOmega^{\mathrm{UL}}$, the voltage safety margin is $M_i = V_i - \underline{V}$, 
with weights set as $\omega_i \propto 1 / M_i$ and normalized to satisfy $\sum_{i}^{n} \omega_i = 1$.

\textbf{W3:} Weights are assigned exclusively to the end-of-feeder nodes (16, 18, 31, and 33), each with 0.15, while all other nodes are assigned 0.05.

\textbf{W4:} Weights are optimized to minimize the PDPL 
using a heuristic approach instead of utilizing the surrogate model.
Specifically, particle swarm optimization is employed.

\textbf{W5:} The proposed WPO framework in this paper.

The weights are applied to the predictive model  
to generate predictions by~\eqref{eq:predict},~\eqref{eq:weightedpredictloss}, and~\eqref{eq:optimalTheta}.
Then the prediction results are applied to the optimization model to 
compute the PDPL by~\eqref{eq:optimization},~\eqref{eq:decisionLoss} and~\eqref{eq:expectedDecisionLoss}. 
Unlike traditional statistical metrics, PDPL evaluates prediction performance by its associated decision quality.
Smaller PDPL indicates superior weight-setting performance. 
The detailed settings (e.g., predictive model structure, hyperparameters, and datasets) of each method are consistent to ensure fair comparison.

\colorblack{
Besides, we also include two existing ``predict-and-optimize'' methods for comparison:}

\colorblack{
\textbf{E2E:} This method adopts neural networks to establish an end-to-end model that directly maps predictive features
to the optimal decisions.}

\colorblack{
\textbf{PE:} This method adopts a linear prediction model and embeds it into the optimization problem, 
as one of the ``Prediction Embedding'' methods.
}

\begin{figure}[htbp]
    \setlength{\abovecaptionskip}{-0.1cm}  
    \setlength{\belowcaptionskip}{-0.1cm}   
    \centering
    \includegraphics[width=0.95\columnwidth]{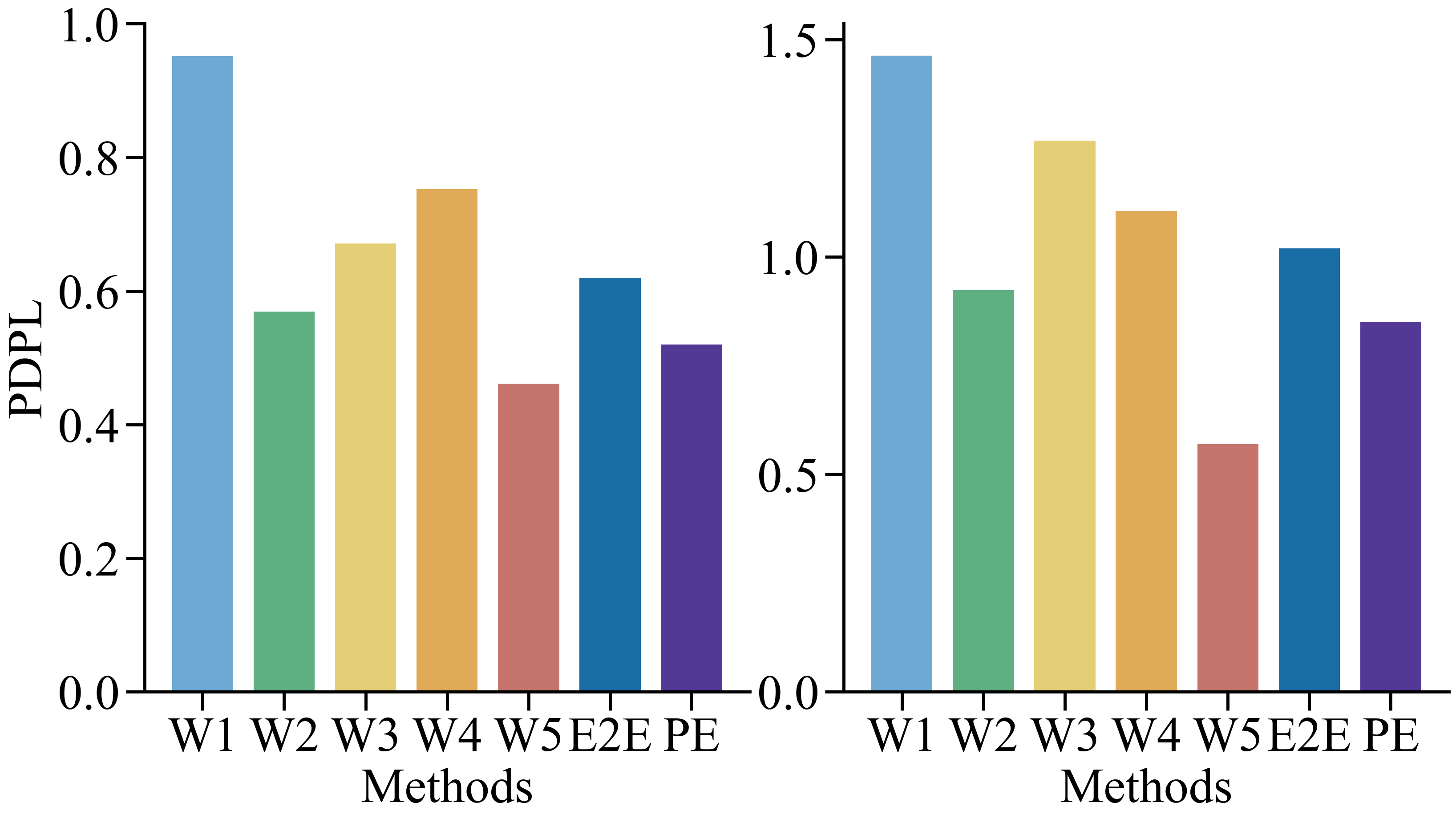}
    \caption{Comparison of different methods: (a) Case 1, (b) Case 2.}
    \label{fig:comparison}
\end{figure}

The comparison results are shown in Fig.~\ref{fig:comparison}.

As shown in Fig.~\ref{fig:comparison},
the proposed WPO framework consistently outperforms other weight-setting methods in both test cases,
achieving the smallest PDPL. 
In terms of the prediction performance,
measured by $\mathbb{E}_{\mathcal{D}^\mathrm{F}}[(\hat{\xi} - \xi)^2]$,
the prediction losses of W1-W5 are: W1 is $3.8 \times 10^{-3}$, W2 is $2.9 \times 10^{-3}$, 
W3 is $4.4 \times 10^{-3}$, W4 is $3.6 \times 10^{-3}$, W5 is $3.3 \times 10^{-3}$, respectively.
The results of prediction error and PDPL across W1-W5 indicate that 
a smaller prediction loss does not necessarily equate to better decision quality. 
This finding further highlights the necessity of implementing an integrated prediction and optimization, 
which aims to optimize decision quality and thereby obtain more valuable prediction outcomes.

W1 represents the traditional predict-then-optimize paradigm with uniform weight-setting method, 
which assigns equal importance to all uncertain variables 
without considering their varying impacts on downstream decision-making. 
This limitation prevents it from addressing the varying importance of uncertain variables in specific optimization problems, 
resulting in relatively high PDPL in certain tasks.

W2 incorporates voltage safety margins of the original network to quantify potential risks. 
While this method captures the general risk distribution of the system, 
its effectiveness depends on the consistency between the risk profile of the original network and the actual risk distribution.
In Case 1, where the risk pattern of the original network aligns closely with the actual risk distribution, 
W2 achieves relatively satisfactory performance. 
However, in Case 2, discrepancies between the original and actual risk distributions, 
induced by load variations, hinder W2's adaptability, resulting in increased decision losses.

W3 assigns weights only to terminal nodes while disregarding the overall voltage risk distribution of the system,
similarly overlooks specific task characteristics, 
thus resulting in suboptimal decision-making outcomes.
W4 suffers from the inherent limitations of heuristic optimization methods,
which largely rely on predefined rules and hyperparameters,
and are susceptible to search space limitations and convergence issues.

In contrast, the proposed weight-setting method W5 achieves the smallest PDPL in both test cases.
By aligning the weight settings with specific risk profiles,
W5 integrates task-specific characteristics directly into the predictive process.
By identifying critical uncertainties that significantly impact downstream optimization tasks
and assigning greater weights to critical uncertainties,
W5 achieves relative more accurate predictions at these nodes,
thus reducing the overall PDPL.
In this way, W5 ensures enhanced prediction-optimization performance,
adaptability across varying risk scenarios,
and interpretability.

\colorblack{
Note that the E2E and PE methods achieve relatively high PDPL than W5 in both test cases.
This is generally because the considered prediction and optimization models are both complex,
so end-to-end learning may not effectively capture the intricate relationships between input features and optimal decisions.
Moreover, the linear prediction model in PE may not adequately provide accurate predictions,
thus leading to suboptimal decisions.
These results further highlight the effectiveness and generality of the proposed WPO framework
in handling complex prediction and optimization tasks.
}

\colorblack{
It has to be mentioned that we do not claim the proposed WPO framework outperforms existing ``predict-and-optimize'' methods.
Rather, WPO framework offers a novel perspective to achieve integrated prediction-and-optimization
by incorporating weights that reflect the varying impacts of uncertainties in specific optimization tasks,
and is more general since it does not rely on specific properties of the prediction and optimization models.
}

\subsection{Scalability Validation}

To validate the scalability of the proposed WPO framework, 
we apply it to a modified IEEE 123-bus distribution network. 
Detailed network configurations and test case settings are provided in~\cite{data}. 
$|\varOmega^{\mathrm{UL}}|$ is increased to 42, 3.5 times that of the 33-bus system.
Compared to the IEEE 33-bus test case, 
the 123-bus network presents a significantly larger system scale with 
increased dimensionality of $\omega$
and more intricate power flow patterns. 
These factors collectively introduce higher complexity to the 
prediction, optimization and surrogate model mapping processes.
Compared with IEEE 33-bus system with 12 uncertain variables,
the parameter count of the MTL model increases from 296,902 to 380,487, 
and the training time increases from 3,571s to 5,939s.
The parameter count of the surrogate model increases from 1,233 to 17,925,
and the training time increases from 63s to 376s.
Though the increased system complexity leads to increased parameter count and training time
in both MTL and surrogate model,
the computational burden remains within an acceptable range,
demonstrating the scalability of the proposed WPO framework.
The comparison results of PDPL of different weight-setting methods
are: W1 is 11.17, W2 is 7.57, W3 is 9.53, W4 is 8.32, W5 is 5.36,
\colorblack{ E2E is 8.85, PE is 9.87.}
This result demonstrates that in a much larger-scale power system,
the proposed WPO framework consistently achieves the lowest PDPL, 
outperforming other methods,
and further validates the scalability and practical applicability of the WPO framework in large-scale power systems.

\section{Conclusion}\label{sec:conclusion}
In this paper, a novel weighted predict-and-optimize framework is proposed for 
uncertainty management in power systems operations.
By introducing weights that quantify the varying impacts of critical uncertainties on specific optimization objectives into the predictive model, 
and optimizing the weights to minimize the problem-driven prediction loss,
WPO achieves integrated and adaptive learning of prediction and optimization.
This mechanism enables WPO to effectively identify and emphasize uncertainties that significantly impact downstream optimization tasks,
and assigns higher weights to these uncertainties to enhance their relative prediction accuracy,
thus reducing the negative impacts of prediction deviations on decision-making outcomes.
As illustrated by extensive case studies on uncertainty management problems in DN,
the presented WPO framework outperforms other weight-setting methods by achieving the smallest PDPL,
and shows strong adaptability, scalability, and interpretability across varying risk scenarios and system scales.

Future work will consider extending the WPO framework to probabilistic prediction and
stochastic optimization problems.

\bibliographystyle{IEEEtran}
\bibliography{IEEEabrv,WPO}

\begin{thebibliography}{10}
\providecommand{\url}[1]{#1}
\csname url@samestyle\endcsname
\providecommand{\newblock}{\relax}
\providecommand{\bibinfo}[2]{#2}
\providecommand{\BIBentrySTDinterwordspacing}{\spaceskip=0pt\relax}
\providecommand{\BIBentryALTinterwordstretchfactor}{4}
\providecommand{\BIBentryALTinterwordspacing}{\spaceskip=\fontdimen2\font plus
\BIBentryALTinterwordstretchfactor\fontdimen3\font minus \fontdimen4\font\relax}
\providecommand{\BIBforeignlanguage}[2]{{%
\expandafter\ifx\csname l@#1\endcsname\relax
\typeout{** WARNING: IEEEtran.bst: No hyphenation pattern has been}%
\typeout{** loaded for the language `#1'. Using the pattern for}%
\typeout{** the default language instead.}%
\else
\language=\csname l@#1\endcsname
\fi
#2}}
\providecommand{\BIBdecl}{\relax}
\BIBdecl
\renewcommand{\BIBentryALTinterwordstretchfactor}{4}

\bibitem{Uncertaintyreview}
L.~A. Roald, D.~Pozo, A.~Papavasiliou \emph{et~al.}, ``Power systems optimization under uncertainty: A review of methods and applications,'' \emph{Electric Power Systems Research}, vol. 214, p. 108725, 2023.

\bibitem{predict1}
M.~Yang, Y.~Huang, C.~Xu \emph{et~al.}, ``Review of several key processes in wind power forecasting: Mathematical formulations, scientific problems, and logical relations,'' \emph{Applied Energy}, vol. 377, p. 124631, 2025.

\bibitem{PTO_UC}
L.~Alvarado-Barrios, Álvaro {Rodríguez del Nozal}, J.~{Boza Valerino} \emph{et~al.}, ``Stochastic unit commitment in microgrids: Influence of the load forecasting error and the availability of energy storage,'' \emph{Renewable Energy}, vol. 146, pp. 2060--2069, 2020.

\bibitem{PTO_reserve}
Y.~Xu, C.~Wan, H.~Liu \emph{et~al.}, ``Probabilistic forecasting-based reserve determination considering multi-temporal uncertainty of renewable energy generation,'' \emph{IEEE Transactions on Power Systems}, vol.~39, no.~1, pp. 1019--1031, 2024.

\bibitem{PTO_EVpredict}
Y.~Zhuang, L.~Cheng, N.~Qi \emph{et~al.}, ``Real-time hosting capacity assessment for electric vehicles: A sequential forecast-then-optimize method,'' \emph{Applied Energy}, vol. 380, p. 125034, 2025.

\bibitem{predicterror}
J.~Wang, Y.~Zhou, Y.~Zhang \emph{et~al.}, ``Risk-averse optimal combining forecasts for renewable energy trading under cvar assessment of forecast errors,'' \emph{IEEE Transactions on Power Systems}, vol.~39, no.~1, pp. 2296--2309, 2024.

\bibitem{physicalpredict}
M.~J. Mayer and G.~Gróf, ``Extensive comparison of physical models for photovoltaic power forecasting,'' \emph{Applied Energy}, vol. 283, p. 116239, 2021.

\bibitem{datapredict}
M.~Sun, T.~Zhang, Y.~Wang \emph{et~al.}, ``Using bayesian deep learning to capture uncertainty for residential net load forecasting,'' \emph{IEEE Transactions on Power Systems}, vol.~35, no.~1, pp. 188--201, 2020.

\bibitem{qi2023}
N.~Qi, P.~Pinson, M.~R. Almassalkhi \emph{et~al.}, ``Chance-constrained generic energy storage operations under decision-dependent uncertainty,'' \emph{IEEE Transactions on Sustainable Energy}, vol.~14, no.~4, pp. 2234--2248, 2023.

\bibitem{DRO}
X.~Shi, Y.~Xu, Q.~Guo \emph{et~al.}, ``Day-ahead distributionally robust optimization-based scheduling for distribution systems with electric vehicles,'' \emph{IEEE Transactions on Smart Grid}, vol.~14, no.~4, pp. 2837--2850, 2023.

\bibitem{DFLreview}
J.~Mandi, J.~Kotary, S.~Berden \emph{et~al.}, ``Decision-focused learning: Foundations, state of the art, benchmark and future opportunities,'' \emph{Journal of Artificial Intelligence Research}, vol.~80, pp. 1623--1701, 2024.

\bibitem{DFLreview1}
R.~Li, H.~Zhang, M.~Sun \emph{et~al.}, ``Decision-oriented learning for future power system decision-making under uncertainty,'' \emph{arXiv preprint arXiv:2401.03680}, 2024.

\bibitem{DFLreview2}
H.~Zhang, R.~Li, Q.~Du \emph{et~al.}, ``Decision-focused learning for power system decision-making under uncertainty,'' \emph{IEEE Transactions on Power Systems}, pp. 1--18, 2025.

\bibitem{PAO_ES}
M.~Yi, S.~Alghumayjan, and B.~Xu, ``Perturbed decision-focused learning for modeling strategic energy storage,'' \emph{IEEE Transactions on Smart Grid}, vol.~16, no.~3, pp. 2574--2586, 2025.

\bibitem{PAO_trading}
T.~Carriere and G.~Kariniotakis, ``An integrated approach for value-oriented energy forecasting and data-driven decision-making application to renewable energy trading,'' \emph{IEEE Transactions on Smart Grid}, vol.~10, no.~6, pp. 6933--6944, 2019.

\bibitem{Beichter}
M.~Beichter, D.~Werling, B.~Heidrich \emph{et~al.}, ``Decision-focused retraining of forecast models for optimization problems in smart energy systems,'' in \emph{Proceedings of the 15th ACM International Conference on Future and Sustainable Energy Systems}, ser. e-Energy '24.\hskip 1em plus 0.5em minus 0.4em\relax New York, NY, USA: Association for Computing Machinery, 2024, p. 170–181.

\bibitem{chen2024neural}
G.~Chen and J.~Qin, ``Neural risk limiting dispatch in power networks: Formulation and generalization guarantees,'' \emph{IEEE Transactions on Power Systems}, pp. 1--13, 2025.

\bibitem{PAO_costinterval}
C.~Zhao, C.~Wan, and Y.~Song, ``Cost-oriented prediction intervals: On bridging the gap between forecasting and decision,'' \emph{IEEE Transactions on Power Systems}, vol.~37, no.~4, pp. 3048--3062, 2022.

\bibitem{PAO_Inertia}
H.~Zhang, R.~Li, Y.~Chen \emph{et~al.}, ``Risk-aware objective-based forecasting in inertia management,'' \emph{IEEE Transactions on Power Systems}, vol.~39, no.~2, pp. 4612--4623, 2024.

\bibitem{PAO_UC1}
X.~Chen, Y.~Yang, Y.~Liu \emph{et~al.}, ``Feature-driven economic improvement for network-constrained unit commitment: A closed-loop predict-and-optimize framework,'' \emph{IEEE Transactions on Power Systems}, vol.~37, no.~4, pp. 3104--3118, 2022.

\bibitem{PAO_VPP1}
Y.~Zhang, M.~Jia, H.~Wen \emph{et~al.}, ``Toward value-oriented renewable energy forecasting: An iterative learning approach,'' \emph{IEEE Transactions on Smart Grid}, vol.~16, no.~2, pp. 1962--1974, 2025.

\bibitem{PAO_Seqmarket}
Y.~Zhang, H.~Wen, Y.~Bian \emph{et~al.}, ``Improving sequential market coordination via value-oriented renewable energy forecasting,'' \emph{IEEE Transactions on Energy Markets, Policy and Regulation}, pp. 1--16, 2025.

\bibitem{chung2022decision}
T.-H. Chung, V.~Rostami, H.~Bastani \emph{et~al.}, ``Decision-aware learning for optimizing health supply chains,'' \emph{arXiv preprint arXiv:2211.08507}, 2022.

\bibitem{PAO_priceES}
L.~Sang, Y.~Xu, H.~Long \emph{et~al.}, ``Electricity price prediction for energy storage system arbitrage: A decision-focused approach,'' \emph{IEEE Transactions on Smart Grid}, vol.~13, no.~4, pp. 2822--2832, 2022.

\bibitem{PAO_voltagecontrol}
L.~Sang, Y.~Xu, H.~Long \emph{et~al.}, ``Safety-aware semi-end-to-end coordinated decision model for voltage regulation in active distribution network,'' \emph{IEEE Transactions on Smart Grid}, vol.~14, no.~3, pp. 1814--1826, 2023.

\bibitem{PAO_costpredict}
J.~Zhang, Y.~Wang, and G.~Hug, ``Cost-oriented load forecasting,'' \emph{Electric Power Systems Research}, vol. 205, p. 107723, 2022.

\bibitem{PAO_wind}
G.~Li and H.-D. Chiang, ``Toward cost-oriented forecasting of wind power generation,'' \emph{IEEE Transactions on Smart Grid}, vol.~9, no.~4, pp. 2508--2517, 2018.

\bibitem{PAO_zhang2023D}
Y.~Zhang, H.~Wen, Y.~Bian \emph{et~al.}, ``Deriving loss function for value-oriented renewable energy forecasting,'' \emph{arXiv preprint arXiv:2310.00571}, 2023.

\bibitem{SPO}
A.~N. Elmachtoub and P.~Grigas, ``Smart ``predict, then optimize'','' \emph{Management Science}, vol.~68, no.~1, pp. 9--26, 2022.

\bibitem{optnet}
B.~Amos and J.~Z. Kolter, ``Optnet: Differentiable optimization as a layer in neural networks,'' in \emph{Proceedings of the 34th International Conference on Machine Learning}, ser. Proceedings of Machine Learning Research, D.~Precup and Y.~W. Teh, Eds., vol.~70.\hskip 1em plus 0.5em minus 0.4em\relax PMLR, 06--11 Aug 2017, pp. 136--145.

\bibitem{PAO_DR}
Y.~Zhang, H.~Wen, T.~Feng \emph{et~al.}, ``Targeted demand response: Formulation, lmp implications, and fast algorithms,'' \emph{arXiv preprint arXiv:2211.14806}, 2022.

\bibitem{weight2}
J.~Wang, C.~Zheng, X.~Yang \emph{et~al.}, ``Enhanceface: Adaptive weighted softmax loss for deep face recognition,'' \emph{IEEE Signal Processing Letters}, vol.~29, pp. 65--69, 2022.

\bibitem{weight3}
Y.~Song, J.~Y.-C. Teoh, K.-S. Choi \emph{et~al.}, ``Dynamic loss weighting for multiorgan segmentation in medical images,'' \emph{IEEE Transactions on Neural Networks and Learning Systems}, vol.~35, no.~8, pp. 10\,651--10\,662, 2024.

\bibitem{gradientconverge}
K.~Scaman, C.~Malherbe, and L.~D. Santos, ``Convergence rates of non-convex stochastic gradient descent under a generic lojasiewicz condition and local smoothness,'' in \emph{Proceedings of the 39th International Conference on Machine Learning}, ser. Proceedings of Machine Learning Research, vol. 162.\hskip 1em plus 0.5em minus 0.4em\relax PMLR, 17--23 Jul 2022, pp. 19\,310--19\,327.

\bibitem{spectralGCN}
T.~N. Kipf and M.~Welling, ``Semi-supervised classification with graph convolutional networks,'' \emph{arXiv preprint arXiv:1609.02907}, 2016.

\bibitem{SOP}
Z.~Wang and S.~Ji, ``Second-order pooling for graph neural networks,'' \emph{IEEE Transactions on Pattern Analysis and Machine Intelligence}, vol.~45, no.~6, pp. 6870--6880, 2023.

\bibitem{MTL_EV}
Y.~Shang, D.~Li, Y.~Li \emph{et~al.}, ``Explainable spatiotemporal multi-task learning for electric vehicle charging demand prediction,'' \emph{Applied Energy}, vol. 384, p. 125460, 2025.

\bibitem{data}
\BIBentryALTinterwordspacing
``Network settings.'' [Online]. Available: \url{https://github.com/Yingrui-Z/Data_for_WPO_Paper}
\BIBentrySTDinterwordspacing

\end{thebibliography}

\end{document}